\documentclass[aps,prd,reprint,superscriptaddress,amsmath,amssymb,nofootinbib]{revtex4-2}

\usepackage{graphicx} 
\usepackage{dcolumn}  
\usepackage{bm}       
\usepackage{hyperref} 
\usepackage{booktabs}
\usepackage{array}
\usepackage{xcolor}
\usepackage{lineno}
\usepackage[T1]{fontenc}
\usepackage{lmodern}   
\usepackage{enumitem}
\usepackage{mathastext}

\begin{document}

\title{First measurement of neutron capture multiplicity in neutrino-oxygen neutral-current quasi-elastic-like interactions using an accelerator neutrino beam}


\newcommand{\INSTHD}{\affiliation{University Autonoma Madrid, Department of Theoretical Physics, 28049 Madrid, Spain}}
\newcommand{\INSTFE}{\affiliation{Boston University, Department of Physics, Boston, Massachusetts, U.S.A.}}
\newcommand{\INSTD}{\affiliation{University of British Columbia, Department of Physics and Astronomy, Vancouver, British Columbia, Canada}}
\newcommand{\INSTGA}{\affiliation{University of California, Irvine, Department of Physics and Astronomy, Irvine, California, U.S.A.}}
\newcommand{\INSTI}{\affiliation{IRFU, CEA, Universit\'e Paris-Saclay, F-91191 Gif-sur-Yvette, France}}
\newcommand{\INSTGB}{\affiliation{University of Colorado at Boulder, Department of Physics, Boulder, Colorado, U.S.A.}}
\newcommand{\INSTFH}{\affiliation{Duke University, Department of Physics, Durham, North Carolina, U.S.A.}}
\newcommand{\INSTJA}{\affiliation{E\"{o}tv\"{o}s Lor\'{a}nd University, Department of Atomic Physics, Budapest, Hungary}}
\newcommand{\INSTEF}{\affiliation{ETH Zurich, Institute for Particle Physics and Astrophysics, Zurich, Switzerland}}
\newcommand{\INSTIG}{\affiliation{VNU University of Science, Vietnam National University, Hanoi, Vietnam}}
\newcommand{\INSTIE}{\affiliation{CERN European Organization for Nuclear Research, CH-1211 Gen\'eve 23, Switzerland}}
\newcommand{\INSTEG}{\affiliation{University of Geneva, Section de Physique, DPNC, Geneva, Switzerland}}
\newcommand{\INSTHJ}{\affiliation{University of Glasgow, School of Physics and Astronomy, Glasgow, United Kingdom}}
\newcommand{\INSTJG}{\affiliation{Ghent University, Department of Physics and Astronomy, Proeftuinstraat 86, B-9000 Gent, Belgium}}
\newcommand{\INSTDG}{\affiliation{H. Niewodniczanski Institute of Nuclear Physics PAN, Cracow, Poland}}
\newcommand{\INSTCB}{\affiliation{High Energy Accelerator Research Organization (KEK), Tsukuba, Ibaraki, Japan}}
\newcommand{\INSTIB}{\affiliation{University of Houston, Department of Physics, Houston, Texas, U.S.A.}}
\newcommand{\INSTED}{\affiliation{Institut de Fisica d'Altes Energies (IFAE) - The Barcelona Institute of Science and Technology, Campus UAB, Bellaterra (Barcelona) Spain}}
\newcommand{\INSTJC}{\affiliation{Institut f\"ur Physik, Johannes Gutenberg-Universit\"at Mainz, Staudingerweg 7, 55128 Mainz, Germany}}
\newcommand{\INSTHH}{\affiliation{Institute For Interdisciplinary Research in Science and Education (IFIRSE), ICISE, Quy Nhon, Vietnam}}
\newcommand{\INSTEI}{\affiliation{Imperial College London, Department of Physics, London, United Kingdom}}
\newcommand{\INSTGF}{\affiliation{INFN Sezione di Bari and Universit\`a e Politecnico di Bari, Dipartimento Interuniversitario di Fisica, Bari, Italy}}
\newcommand{\INSTBE}{\affiliation{INFN Sezione di Napoli and Universit\`a di Napoli, Dipartimento di Fisica, Napoli, Italy}}
\newcommand{\INSTBF}{\affiliation{INFN Sezione di Padova and Universit\`a di Padova, Dipartimento di Fisica, Padova, Italy}}
\newcommand{\INSTBD}{\affiliation{INFN Sezione di Roma and Universit\`a di Roma ``La Sapienza'', Roma, Italy}}
\newcommand{\INSTEB}{\affiliation{Institute for Nuclear Research of the Russian Academy of Sciences, Moscow, Russia}}
\newcommand{\INSTHI}{\affiliation{International Centre of Physics, Institute of Physics (IOP), Vietnam Academy of Science and Technology (VAST), 10 Dao Tan, Ba Dinh, Hanoi, Vietnam}}
\newcommand{\INSTJD}{\affiliation{ILANCE, CNRS – University of Tokyo International Research Laboratory, Kashiwa, Chiba 277-8582, Japan}}
\newcommand{\INSTHA}{\affiliation{Kavli Institute for the Physics and Mathematics of the Universe (WPI), The University of Tokyo Institutes for Advanced Study, University of Tokyo, Kashiwa, Chiba, Japan}}
\newcommand{\INSTID}{\affiliation{Keio University, Department of Physics, Kanagawa, Japan}}
\newcommand{\INSTIF}{\affiliation{King's College London, Department of Physics, Strand, London WC2R 2LS, United Kingdom}}
\newcommand{\INSTCC}{\affiliation{Kobe University, Kobe, Japan}}
\newcommand{\INSTCD}{\affiliation{Kyoto University, Department of Physics, Kyoto, Japan}}
\newcommand{\INSTEJ}{\affiliation{Lancaster University, Physics Department, Lancaster, United Kingdom}}
\newcommand{\INSTII}{\affiliation{Lawrence Berkeley National Laboratory, Berkeley, California, U.S.A.}}
\newcommand{\INSTBA}{\affiliation{Ecole Polytechnique, IN2P3-CNRS, Laboratoire Leprince-Ringuet, Palaiseau, France}}
\newcommand{\INSTFC}{\affiliation{University of Liverpool, Department of Physics, Liverpool, United Kingdom}}
\newcommand{\INSTFI}{\affiliation{Louisiana State University, Department of Physics and Astronomy, Baton Rouge, Louisiana, U.S.A.}}
\newcommand{\INSTIH}{\affiliation{Joint Institute for Nuclear Research, Dubna, Moscow Region, Russia}}
\newcommand{\INSTHB}{\affiliation{Michigan State University, Department of Physics and Astronomy,  East Lansing, Michigan, U.S.A.}}
\newcommand{\INSTCE}{\affiliation{Miyagi University of Education, Department of Physics, Sendai, Japan}}
\newcommand{\INSTDF}{\affiliation{National Centre for Nuclear Research, Warsaw, Poland}}
\newcommand{\INSTFJ}{\affiliation{State University of New York at Stony Brook, Department of Physics and Astronomy, Stony Brook, New York, U.S.A.}}
\newcommand{\INSTEH}{\affiliation{STFC, Rutherford Appleton Laboratory, Harwell Oxford,  and  Daresbury Laboratory, Warrington, United Kingdom}}
\newcommand{\INSTGJ}{\affiliation{Okayama University, Department of Physics, Okayama, Japan}}
\newcommand{\INSTCF}{\affiliation{Osaka Metropolitan University, Department of Physics, Osaka, Japan}}
\newcommand{\INSTGG}{\affiliation{Oxford University, Department of Physics, Oxford, United Kingdom}}
\newcommand{\INSTIC}{\affiliation{University of Pennsylvania, Department of Physics and Astronomy,  Philadelphia, Pennsylvania, U.S.A.}}
\newcommand{\INSTGC}{\affiliation{University of Pittsburgh, Department of Physics and Astronomy, Pittsburgh, Pennsylvania, U.S.A.}}
\newcommand{\INSTGD}{\affiliation{University of Rochester, Department of Physics and Astronomy, Rochester, New York, U.S.A.}}
\newcommand{\INSTHC}{\affiliation{Royal Holloway University of London, Department of Physics, Egham, Surrey, United Kingdom}}
\newcommand{\INSTBC}{\affiliation{RWTH Aachen University, III. Physikalisches Institut, Aachen, Germany}}
\newcommand{\INSTJF}{\affiliation{School of Physics and Astronomy, University of Minnesota, Minneapolis, Minnesota, U.S.A.}}
\newcommand{\INSTJB}{\affiliation{Departamento de F\'isica At\'omica, Molecular y Nuclear, Universidad de Sevilla, 41080 Sevilla, Spain}}
\newcommand{\INSTFB}{\affiliation{University of Sheffield, School of Mathematical and Physical Sciences, Sheffield, United Kingdom}}
\newcommand{\INSTDI}{\affiliation{University of Silesia, Institute of Physics, Katowice, Poland}}
\newcommand{\INSTIA}{\affiliation{SLAC National Accelerator Laboratory, Stanford University, Menlo Park, California, U.S.A.}}
\newcommand{\INSTBB}{\affiliation{Sorbonne Universit\'e, CNRS/IN2P3, Laboratoire de Physique Nucl\'eaire et de Hautes Energies (LPNHE), Paris, France}}
\newcommand{\INSTJE}{\affiliation{South Dakota School of Mines and Technology, 501 East Saint Joseph Street, Rapid City, SD 57701, United States}}
\newcommand{\INSTCH}{\affiliation{University of Tokyo, Department of Physics, Tokyo, Japan}}
\newcommand{\INSTBJ}{\affiliation{University of Tokyo, Institute for Cosmic Ray Research, Kamioka Observatory, Kamioka, Japan}}
\newcommand{\INSTCG}{\affiliation{University of Tokyo, Institute for Cosmic Ray Research, Research Center for Cosmic Neutrinos, Kashiwa, Japan}}
\newcommand{\INSTHF}{\affiliation{Institute of Science Tokyo, Department of Physics, Tokyo}}
\newcommand{\INSTGI}{\affiliation{Tokyo Metropolitan University, Department of Physics, Tokyo, Japan}}
\newcommand{\INSTHG}{\affiliation{Tokyo University of Science, Faculty of Science and Technology, Department of Physics, Noda, Chiba, Japan}}
\newcommand{\INSTB}{\affiliation{TRIUMF, Vancouver, British Columbia, Canada}}
\newcommand{\INSTJH}{\affiliation{University of Toyama, Department of Physics, Toyama, Japan}}
\newcommand{\INSTDJ}{\affiliation{University of Warsaw, Faculty of Physics, Warsaw, Poland}}
\newcommand{\INSTDH}{\affiliation{Warsaw University of Technology, Institute of Radioelectronics and Multimedia Technology, Warsaw, Poland}}
\newcommand{\INSTIJ}{\affiliation{Tohoku University, Faculty of Science, Department of Physics, Miyagi, Japan}}
\newcommand{\INSTFD}{\affiliation{University of Warwick, Department of Physics, Coventry, United Kingdom}}
\newcommand{\INSTEA}{\affiliation{Wroclaw University, Faculty of Physics and Astronomy, Wroclaw, Poland}}
\newcommand{\INSTHE}{\affiliation{Yokohama National University, Department of Physics, Yokohama, Japan}}
\newcommand{\INSTH}{\affiliation{York University, Department of Physics and Astronomy, Toronto, Ontario, Canada}}

\INSTHD
\INSTFE
\INSTD
\INSTGA
\INSTI
\INSTGB
\INSTFH
\INSTJA
\INSTEF
\INSTIG
\INSTIE
\INSTEG
\INSTHJ
\INSTJG
\INSTDG
\INSTCB
\INSTIB
\INSTED
\INSTJC
\INSTHH
\INSTEI
\INSTGF
\INSTBE
\INSTBF
\INSTBD
\INSTEB
\INSTHI
\INSTJD
\INSTHA
\INSTID
\INSTIF
\INSTCC
\INSTCD
\INSTEJ
\INSTII
\INSTBA
\INSTFC
\INSTFI
\INSTIH
\INSTHB
\INSTCE
\INSTDF
\INSTFJ
\INSTEH
\INSTGJ
\INSTCF
\INSTGG
\INSTIC
\INSTGC
\INSTGD
\INSTHC
\INSTBC
\INSTJF
\INSTJB
\INSTFB
\INSTDI
\INSTIA
\INSTBB
\INSTJE
\INSTCH
\INSTBJ
\INSTCG
\INSTHF
\INSTGI
\INSTHG
\INSTB
\INSTJH
\INSTDJ
\INSTDH
\INSTIJ
\INSTFD
\INSTEA
\INSTHE
\INSTH

\author{K.\,Abe}\INSTBJ
\author{S.\,Abe}\INSTBJ
\author{R.\,Akutsu}\INSTCB
\author{H.\,Alarakia-Charles}\INSTEJ
\author{Y.I.\,Alj Hakim}\INSTFB
\author{S.\,Alonso Monsalve}\INSTEF
\author{L.\,Anthony}\INSTEI
\author{S.\,Aoki}\INSTCC
\author{K.A.\,Apte}\INSTEI
\author{T.\,Arai}\INSTCH
% ...

\collaboration{The T2K Collaboration}\noaffiliation

\date{\today}

\begin{abstract}
We report the first measurement of neutron capture multiplicity in neutrino-oxygen neutral-current quasi-elastic-like interactions at the gadolinium-loaded Super-Kamiokande detector using the T2K neutrino beam, which has a peak energy of about 0.6 GeV. 
A total of 30 neutral-current quasi-elastic-like event candidates were selected from T2K data corresponding to an exposure of \(1.76 \times 10^{20}\) protons on target. The $\gamma$ ray signals resulting from neutron captures were identified using a neural network. The flux-averaged mean neutron capture multiplicity was measured to be \(1.37 \pm 0.33\)~(stat.)~\(^{+0.17}_{-0.27}\)~(syst.), which is compatible within 2.3~sigma than predictions obtained using our nominal simulation.
We discuss potential sources of systematic uncertainty in the prediction and demonstrate that a significant portion of this discrepancy arises from the modeling of hadron-nucleus interactions in the detector medium. 

\end{abstract}
\maketitle

\section{\label{sec:1_Intro}Introduction}

Precise measurements of neutral-current quasi-elastic (NCQE) interactions on oxygen, the dominant neutral-current process at neutrino energies of approximately 0.1~-~1~GeV, are crucial for searches involving rare processes in water Cherenkov detectors. 
For example, searches for the diffuse supernova neutrino background (DSNB)~\cite{DSNB_SK4, DSNB_SK6} face significant background contributions from NCQE interactions of atmospheric neutrinos, which are affected by large uncertainties. 
The NCQE interactions between $\nu$($\bar{\nu}$) and $^{16}\text{O}$ can be expressed~\cite{NCQE_Ankowski} as follows:
\begin{equation}
    \begin{aligned}
        \nu(\bar{\nu}) + \textnormal{$^{16}$O} & \rightarrow \nu(\bar{\nu}) + \textnormal{$^{15}$O} + \gamma + {\it n} \\
        \nu(\bar{\nu}) + \textnormal{$^{16}$O} & \rightarrow \nu(\bar{\nu}) + \textnormal{$^{15}$N} + \gamma + {\it p}. 
    \end{aligned}
    \label{eq:NCQE_Eq}
\end{equation}
The recoil nucleon is subject to final state interactions as it propagates through the initial target nucleus, which can alter the particle's kinematics, generate additional recoil particles, or induce nuclear de-excitation processes that emit $\gamma$ rays.
Additionally, nucleons escaping the nucleus may undergo further nucleon-nucleus interactions as they traverse the detector medium, collectively labelled as secondary interactions.
Following such interactions, the excited nucleus promptly de-excites to its ground state, emitting secondary $\gamma$ rays in addition to the initial (primary) $\gamma$, as well as neutrons or other particles. 
These primary and secondary $\gamma$ rays can be used to identify NCQE interactions and were used in a previous T2K publication~\cite{NCQE_T2K_Run1-9} to measure the flux-averaged NCQE interaction cross section. 
Recoil hadrons were not measured in that study, though neutrons can be identified via de-excitation $\gamma$ rays emitted when they capture on detector nuclei.


Precise knowledge of the neutron capture multiplicity is critical for estimation of background to the inverse beta decay signal of DSNB events, which are expected to have exactly one neutron, from atmospheric NCQE events, which are often accompanied by multiple neutrons.
While previous measurements have been made with atmospheric neutrinos by SK~\cite{NCQE_SKIV, NCQE_SKVI}, they suffer large uncertainties from the flux model and from 
lower signal purity due to contamination of non-NCQE interactions as well as accidental-coincidence backgrounds.
Those studies also indicate that secondary interactions have a large impact on the number of neutrons populating the final state in Monte Carlo (MC) simulations.
The T2K neutrino beam, with an energy range similar to that of atmospheric neutrino backgrounds to DSNB searches, offers a nearly ideal means of overcoming these limitations. 
Not only does the beam timing allow for precise selection of signal events and for background rejection, but the tightly constrained flux at T2K allows for smaller uncertainties on both the measurements of NCQE interaction cross section and the neutron capture multiplicity.





This paper reports the first neutron capture multiplicity measurement of neutral-current quasi-elastic-like (NCQE-like) interactions with the T2K neutrino beam. 
In the following, the term “NCQE-like” is used inclusively to denote the topology listed in Eq. (\ref{eq:NCQE_Eq}) as well as neutral-current two-particle two-hole (NC 2p2h) interactions. NC 2p2h can eject an extra $np$, $nn$, or $pp$ pair, adding recoil nucleons that mimic the NCQE signal~\cite{NCQE_T2K_Run1-9}.  

This article is organized as follows. 
The experiment is described in Sec.~II while Sec.~III describes the event simulation. 
Details of the event reconstruction and selection methods, including the neutron detection method, are given in Sec.~IV. 
Finally, the results and discussion are given in Secs.~V and VI, before concluding in Sec.~VII.

\section{\label{sec:2_T2K}The T2K Experiment}

The T2K experiment~\cite{T2K_Exp} is a long-baseline accelerator neutrino experiment using the J-PARC neutrino beam as well as a suite of near and far detectors for its measurements. 
Its primary physics program consists of precision measurements of the neutrino oscillation parameters and detailed studies of neutrino interactions at both its near and far detectors.

While T2K has operated with different configurations since the start of operations in 2009, this study focuses on data taken between March 9, 2021, and April 27, 2021, known as Run 11 and corresponding to 1.76~$\times$~10$^{20}$ protons on target (POT), for which the following beam settings apply.
The neutrino beam production begins with protons grouped into eight bunches per spill with each bunch separated by approximately 581~ns.
The J-PARC Main Ring synchrotron accelerates these bunches to 30~GeV/c and directs them to a graphite target in the neutrino beamline with a 2.48-second repetition rate~\cite{T2K_Flux_2013}. 
Hadrons emerging from the proton-target interaction, such as pions and kaons, are focused along the proton beam direction and charge-selected by three magnetic horns~\cite{T2K_Maghorn} operated at $\pm 250$~kA. 
Pions and kaons decay within a dedicated 96~m decay volume located downstream of the magnetic horns, producing neutrinos along the beamline direction.
During Run 11 the horns were operated at $+250$~kA, which focuses positively charged hadrons and produces a predominantly muon neutrino beam from their decays.
At the end of the decay volume, a beam dump, a muon monitor~\cite{T2K_MUMON}, and WAGASCI-BabyMIND~\cite{T2K_BabyMIND} are positioned to indirectly monitor the neutrino beam's direction, width, and yield. 
The near detector complex, consisting of the INGRID~\cite{T2K_INGRID} and ND280~\cite{T2K_ND280_1,T2K_ND280_2} detectors, is located 280 meters downstream of the graphite target.
INGRID is placed on the neutrino beam axis and is responsible for monitoring the direction and intensity of the neutrino beam.
ND280, on the other hand, is located 2.5$^\circ$~off-axis and measures the neutrino energy spectrum and the neutrino flavor composition of the beam before neutrino oscillation effects become significant.

The Super-Kamiokande (SK) detector ~\cite{SK_2003, Gd_2022} serves as T2K's far detector, used to study neutrino oscillations and, in this analysis, the neutrino neutral-current interactions. 
SK is located 295~kilometers downstream and sits 2.5$^\circ$~off-axis with respect to the proton direction. 
It is a cylindrical 50~kiloton water Cherenkov detector consisting of an inner detector (ID) and an outer detector (OD) that is located 1~km underground.  
The ID, which measures 33.8~meters in diameter and 36.2~meters in height, is equipped with 11,129~inward-facing photomultiplier tubes (PMT), each 20~inches in diameter. 
Functioning primarily as a veto, the OD is a 2-meter-thick cylindrical shell surrounding the ID, which is viewed by 
1,885 8-inch outward-facing PMTs.
The OD is separated from the ID by an uninstrumented region 55 cm in width.

Beam timing information is synchronized between J-PARC and SK using a GNSS system. 
At SK, a dedicated event trigger is issued corresponding to the beam arrival time, and it initiates recording of all PMT hit charges and times in a time period of $500~\mu\text{s}$ before and after the trigger.
In 2020, 13~tons of $Gd_2(SO_4)_3 \cdot 8H_2O$ were dissolved into SK, leading to a Gd concentration of 0.011\%(becomes 0.01\% later), for the purpose of enhancing SK's neutron detection capability.
For this reason, the Run 11 data set has significantly enhanced neutron detection capabilities compared to the earlier pure water data taken at T2K. 
Further details on this upgrade to SK are provided in Ref.~\cite{Gd_2022}.

\section{\label{sec:3_Sim}Event Simulation}

This study employs a multi-stage simulation in the T2K experiment, simulating the neutrino flux, neutrino-nucleus interactions, and the detector response.

\subsection{Neutrino Flux}
\sloppy
The neutrino beam flux calculation uses simulations incorporating \textsc{fluka}~\cite{FLUKA} and \textsc{geant}\scalebox{0.8}{3}~\cite{T2K_G3flux}. These packages simulate hadronic interactions, particle transport, and particle decay within the neutrino beamline. 
The production cross sections for pions and kaons are renormalized based on data from the NA61/SHINE experiment, which utilizes a replica of the T2K targets~\cite{T2K_NA61_2019} as well as thin targets~\cite{T2K_NA61_2011, T2K_NA61_2012, T2K_NA61_2016}.
Using NA61/SHINE 2009 replica-target data, the new flux tuning~\cite{T2K_Flux_2023} lowers the peak-region uncertainty from 9--12\% to 5--8\% for the primary $\nu_{\mu}$ fluxes, while the wrong-sign components ($\bar{\nu}_{\mu}$) remain limited by off-target interactions to 6--8\%.

The flux peaks at approximately 0.6~GeV with a width of a few hundred MeV and consists predominantly of $\nu_\mu$ ($\sim$92.4\%) and $\bar{\nu}_\mu$($\sim$6.4\%).

\begin{figure}[htbp]
\centering
\includegraphics[width=0.4\textwidth]{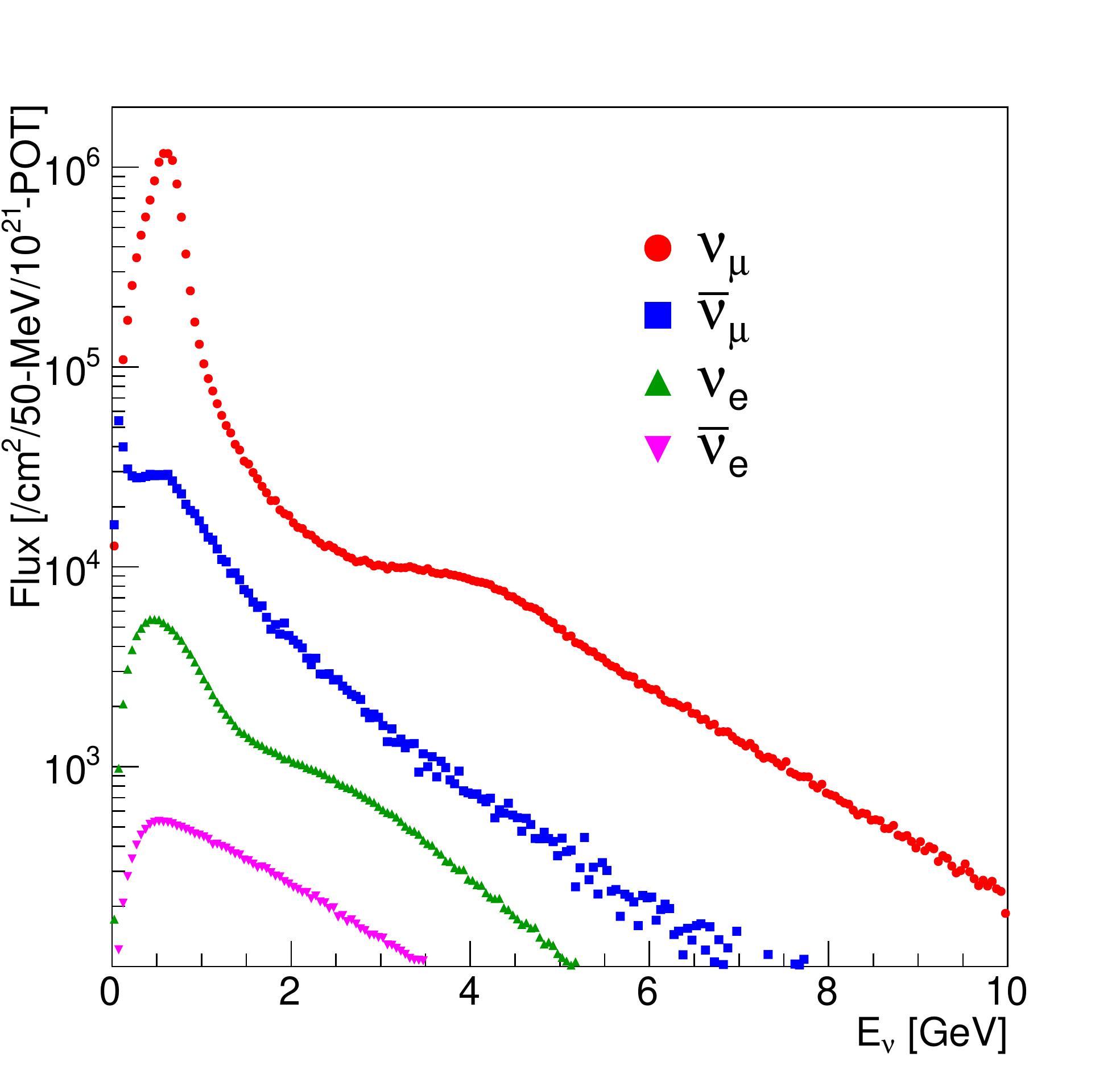}
\caption{The predicted neutrino flux for T2K Run~11 at SK without neutrino oscillations.}


\label{fig:Run11Flux}
\end{figure}

Fig.~\ref{fig:Run11Flux} illustrates the predicted neutrino flux at SK in the absence of neutrino oscillations
for the data set used in this analysis.


\begin{figure*}[htbp]
\centering
\includegraphics[width=\textwidth]{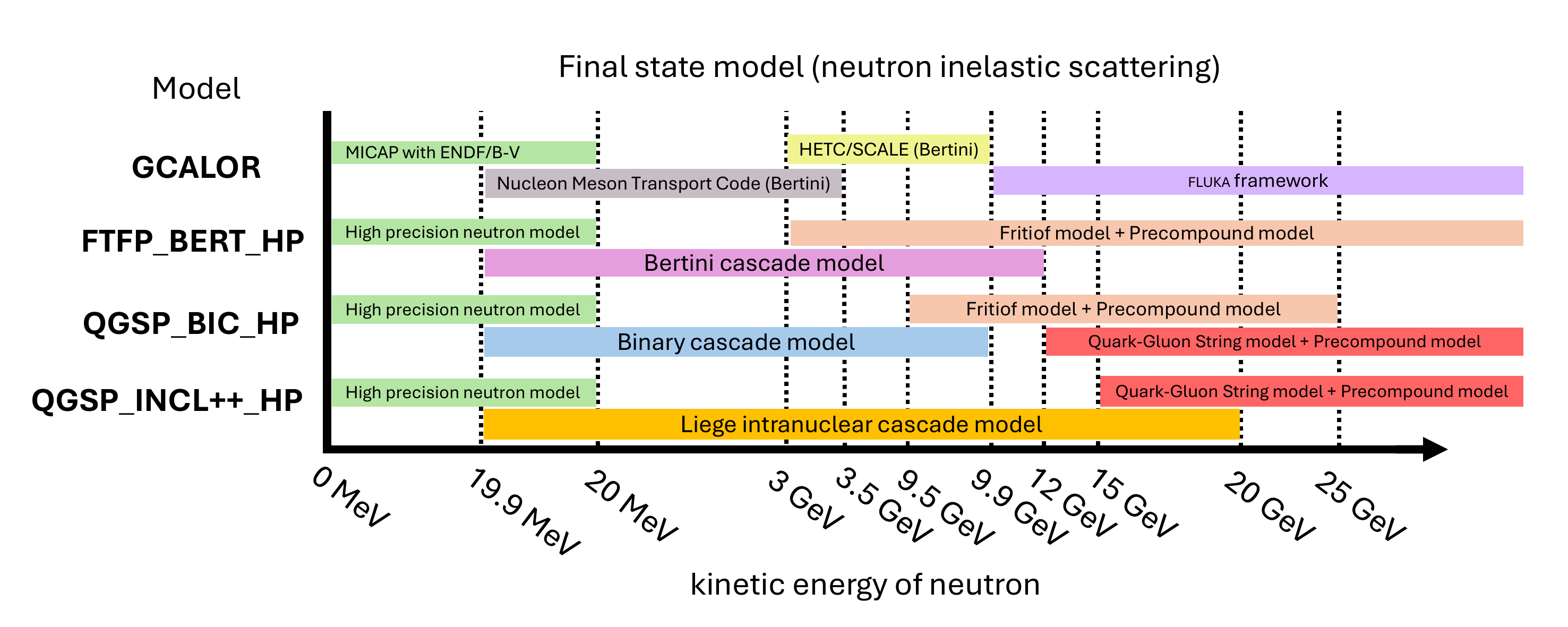}
\caption{Illustration of various “physics lists” used for neutron inelastic scattering in SKDETSIM and SKG4. The horizontal axis indicates the kinetic energy of the incoming neutron. A physics list in \textsc{geant}\scalebox{0.8}{4} defines which interaction models and cross sections are used over different energy ranges. In this study, SKDETSIM (based on \textsc{geant}\scalebox{0.8}{3}) employs GCALOR~\cite{GCALOR}, while SKG4 (based on \textsc{geant}\scalebox{0.8}{4}) can choose from multiple physics lists~\cite{GEANT4,Sakai_thesis}. Examples include FTFP\_BERT\_HP (Fritiof Parton \& Precompound + Bertini Cascade + High Precision neutron model), QGSP\_BIC\_HP (Quark-Gluon String \& Precompound + BInary Cascade + High Precision neutron model), and INCL++\_HP (Liège INtranuclear Cascade + High Precision neutron model). 
}
\label{fig:SITable}
\end{figure*}


\subsection{Neutrino Interaction}
\label{sec:neutrino_interaction}
The NEUT MC code(version 5.6.3)~\cite{NEUT563, NEUT_detail} is used to simulate neutrino-nucleon interactions as well as subsequent intranuclear reactions (henceforth referred to as ``final state interactions'' or FSI) between outgoing hadrons and the $^{16}\text{O}$ nuclear medium.
These interactions can result in altered hadron kinematics, particle absorption, or the production of new hadrons, making the hadrons observed outside the nucleus different from those initially produced. 
NEUT uses the Bertini intranuclear cascade model~\cite{NEUT_INC} to propagate nucleons in a semi-classical manner through the nucleus, calculating interaction probabilities at each propagation step based on the Woods-Saxon nuclear density profile~\cite{Woods_Saxon}. 
For NCQE and charged-current quasi-elastic (CCQE) interactions, the nucleon momentum distribution is based on the spectral function by Benhar et al.~\cite{NCQE_Benhar, NEUT_Abe}. 
This study employs the NEUT 5.6.3 default settings, using a dipole form factor with the axial-vector mass set to $\textit{M}_A^{\text{QE}} = 1.21$~GeV/$c^2$, the strange axial coupling constant ($\textit{g}_A^s$) set to 0.0, and the Fermi momentum for oxygen is set to 209~MeV/$c$. 
NEUT incorporates BBBA05 vector form factors~\cite{BBBA05}. 
The Valencia model by Nieves et al.~\cite{2p2h} is used to simulate charged-current two-particle two-hole (2p2h) interactions, while NC 2p2h interactions are not included. 
Single pion production is simulated using the Rein and Sehgal model~\cite{Rein_Sehgal, Berger_Sehgal, Graczyk_Sobczyk}, with the axial-vector mass set to $\textit{M}_A^{\text{RES}} = 0.95$ \text{GeV}/$c^2$. 
Deep inelastic scattering is simulated based on the GRV98 parton distribution~\cite{GRV98}, with additional corrections from Bodek and Yang~\cite{Bodek_Yang}. 
The neutrino interaction parameters used in NEUT for this study are summarized in Table~\ref{tab:NEUT}.

\begin{table*}[htbp]
\centering
\caption{Model configuration in NEUT 5.6.3.}
\label{tab:NEUT}
\scalebox{1.00}{
\begin{tabular}{@{}llll@{}}
\toprule
Channel       & Model          &   Parameter     & Value or model \\ \midrule
NCQE          & Ankowski et al.\cite{NCQE_Ankowski} & $\textit{M}^{QE}_A$                  & 1.21 GeV \\
              &                            & $\textit{g}^{s}_A$              & 0.0 \\
              &                            & Fermi momentum         & 209 MeV/c \\
              &                            & VFF                    & BBBA05\cite{BBBA05}\\
              &                            & Nuclear model          & Spectral Function\cite{NCQE_Benhar, NEUT_Abe}\\
CCQE          & Ankowski et al.\cite{NCQE_Ankowski} &  Nuclear model          & Spectral Function\cite{NCQE_Benhar, NEUT_Abe}\\
\hline 
CC 2p2h       & Nieves et al.\cite{2p2h}   &                        & \\
CC and NC RES & Rein and Sehgal~\cite{Rein_Sehgal}  & FF                     & Dipole form \\
              &                            & $\textit{M}^{RES}_A$                  & 0.95 GeV \\
CC and NC DIS & GRV98 PDF\cite{GRV98}      & \multicolumn{2}{l}{with modification by Bodek and Yang\cite{Bodek_Yang}} \\
CC and NC COH & Berger-Sehgal\cite{Berger_Sehgal} &                        & \\
\hline 
FSI           & NEUT cascade               & Nucleon cross section  & Bertini et al.\cite{NEUT_INC} \\
\bottomrule
\end{tabular}
}
\end{table*}


Nuclear de-excitation is also simulated by NEUT. 
After an initial neutrino-nucleus interaction, the excited state of the nucleus in Eq.~(\ref{eq:NCQE_Eq}) is determined based on probabilities outlined in Ref.~\cite{NCQE_Ankowski}, with possible states including $({\it p}_{1/2})^{-1}$, $({\it p}_{3/2})^{-1}$, $({\it s}_{1/2})^{-1}$ and others. 
The respective production probabilities are 0.158, 0.3515, 0.1055, and 0.385.
The decay products of each state and their branching ratios are summarized in Ref.~\cite{Sakai_thesis}, based on $\gamma$ ray branching ratio measurements reported in Refs.\cite{BRatio_Exp, BRatio_Exp2, BRatio_Exp3}. 
The \textit{others} category, for which $\gamma$ ray emission measurements are lacking, is treated identically to the $({\it s}_{1/2})^{-1}$ simulations. 

\subsection{Detector Simulation}
Interactions between particles and water to produce and propagate Cherenkov photons are modelled by the SK detector simulation as well as the subsequent response of hit PMTs.


Two simulation packages are employed in our study: SKDETSIM~\cite{T2K_Exp}, which is based on \textsc{geant}\scalebox{0.8}{3} (\textsc{geant}\scalebox{0.8}{3.2.1})~\cite{GEANT3}, and SKG4~\cite{SKG4}, which is built using \textsc{geant}\scalebox{0.8}{4} (\textsc{geant}\scalebox{0.8}{4.10.5p01})~\cite{GEANT4}. 
This study's primary MC is based on SKDETSIM due to its consistent use throughout the history of both SK and T2K.
SKG4 is additionally adopted to make use of the extensive array of hadron-nucleus interaction models 
(termed ``secondary interaction'' or SI models below).
SKDETSIM and SKG4 are individually calibrated to accurately model the transport of Cherenkov photons and the PMT response. 
This includes the calibration of the optical scattering processes, photon absorption, the asymmetry of the water quality between the top and bottom regions of the detector,  the reflection of light by detector materials, as well as the charge and timing response of the PMTs.
Additional details are provided in Ref.~\cite{SK_Calibration}. 

%

Since neutrons produced through NCQE interactions typically have kinetic energies around $\mathcal{O}(100)$ MeV at T2K,
neutron-nucleus interactions can significantly affect the observed result.
The choice of SI model can therefore create notable differences in the total number of predicted $\gamma$ rays and neutrons.
SKDETSIM is the primary choice in this study, which adopts the GCALOR~\cite{GCALOR} model. 
However, for neutron energies below 20~MeV MICAP~\cite{MICAP} is employed, which models 
neutrons, nuclei, and neutron capture by hydrogen based on experimental cross section data from the ENDF/B-V library~\cite{ENDF}.
For nucleon propagation above 20~MeV, GCALOR invokes NMTC~\cite{NMTC}, which handles nucleons up to 3.5~GeV and charged pions up to 2.5~GeV. 
At energies just exceeding these thresholds, the HETC/SCALE approach~\cite{HETC_SCALE}, which is based on the Bertini cascade, is used. 
Finally, \textsc{fluka}~\cite{FLUKA} is applied when energies exceed 10~GeV. 

SKG4 is employed in order to evaluate the impact of different SI models using the ``physics list'' mechanism provided in \textsc{geant}\scalebox{0.8}{4}~\cite{GEANT4}. 
Three physics lists are chosen in this study: {\tt FTFP\_BERT\_HP} (BERT), {\tt QGSP\_INCL++\_HP} (INCL++), and {\tt QGSP\_BIC\_HP} (BIC). 
The Bertini cascade model (BERT)~\cite{BERT} has been widely used in \textsc{geant}\scalebox{0.8}{3} and \textsc{geant}\scalebox{0.8}{4} simulations. 
The Liège Intranuclear cascade model (INCL++)~\cite{INCL_2013} and the Binary Cascade model (BIC) use the binary cascade approach~\cite{BIC}. 
The primary distinction among BERT, INCL++, and BIC lies in their criteria for terminating the intranuclear cascade process and their selection of de-excitation model.
While the BERT model is known for its relatively simple and highly parameterized pre-equilibrium and nuclear de-excitation models~\cite{BERT}, INCL++ and BIC adopt a more detailed nuclear de-excitation model, \texttt{G4PreCompoundModel}~\cite{G4PreCompoundModel}, which offers better precision. 
These differences lead to varied neutron capture multiplicity predictions as is demonstrated below.
Fig.~\ref{fig:SITable} provides an overview of the SI interaction models used in this study.

The SK upgrade~\cite{Gd_2022} enhances neutron detection efficiency by taking advantage of Gd's large neutron capture cross section and the emission of $\gamma$ rays totaling approximately 8~MeV in energy during de-excitation.
The natural abundance of $^{157}$Gd and $^{155}$Gd are 15.65\% and 14.80\%~\cite{GdComp} and at a neutron energy of approximately 0.0253~eV their capture cross sections are 254,000 and 60,900~barns, respectively~\cite{GdXsec}. 
These can be compared with hydrogen's cross section of 0.33~barns~\cite{GdXsec}. 
Consequently, with a Gd concentration around 0.011\%, approximately 52\% of neutrons will be captured on Gd nuclei. 
This study adopts the ANNRI-Gd model~\cite{ANNRIGd} from \textsc{geant}\scalebox{0.8}{4} to model $\gamma$ ray emission from 
Gd nuclear de-excitation following neutron capture.



\section{\label{sec:4_Reconst}Event Reconstruction and Selection}

\begin{figure}[htbp]
\centering
\includegraphics[width=0.5\textwidth]{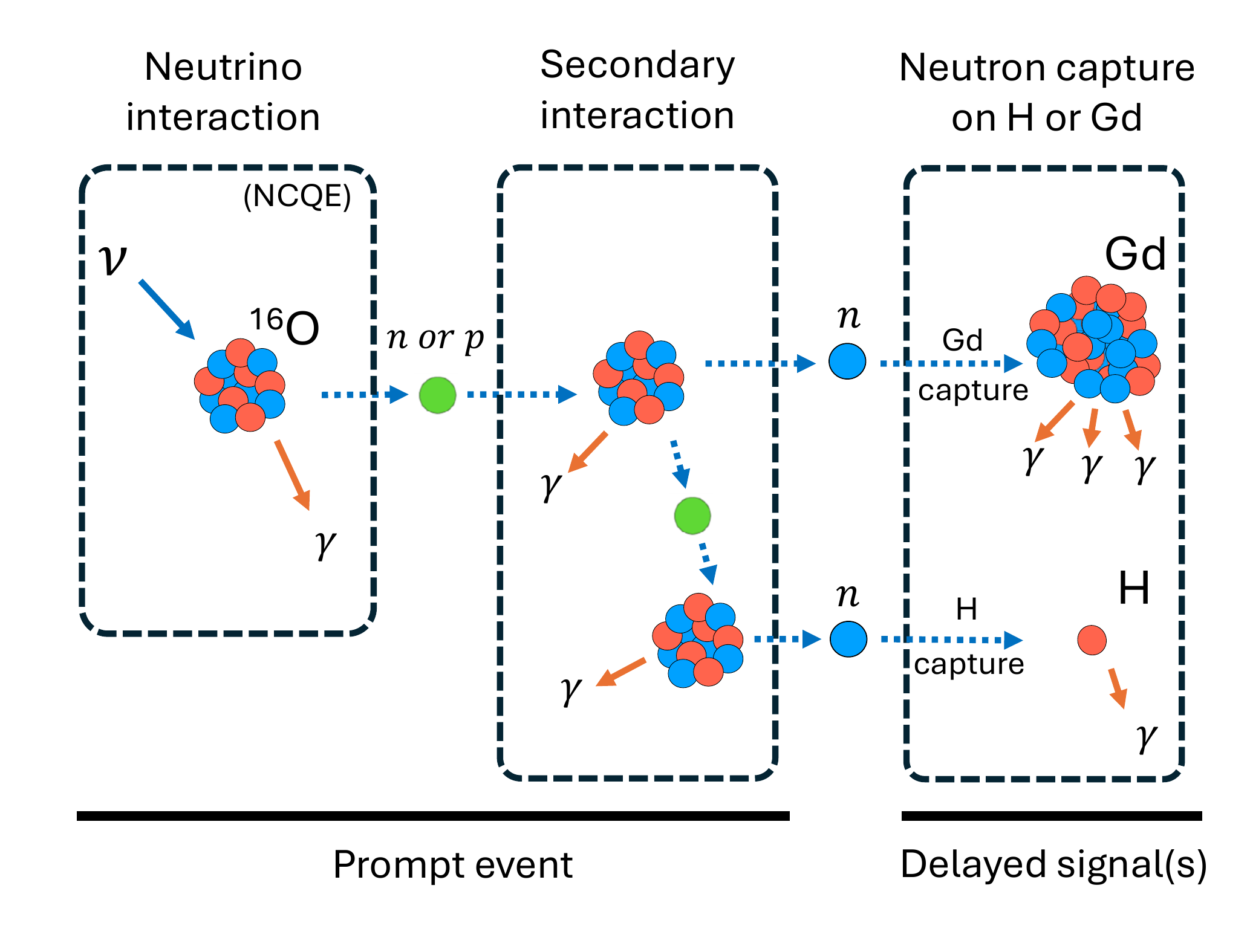}
\caption{Schematic of the NCQE interaction. The $\gamma$ emission from the primary neutrino interaction and the following final state interaction, including secondary nuclear interactions, marks the ``prompt'' event, while the produced neutrons are captured by hydrogen or gadolinium, accompanied by the delayed $\gamma$ emission, representing the ``delayed'' signal.}
\label{fig:NCQEdemo}
\end{figure}

NCQE-like events are characterized by a delayed pair of triggers in the detector, one occurring promptly after the expected arrival time of the neutrino beam, followed by one or more delayed triggers within the same beam spill window, with spills arriving every 2.48 seconds. 
Triggers are issued to record all the PMT charge and timing information in a time period of 500~$\mu$s before and after the beam arrival time in this analysis.
Fig.~\ref{fig:NCQEdemo} illustrates the target topology.
The prompt event consists of one or more $\gamma$ rays produced by the de-excitation of nuclear remnant produced by the $\nu$-$^{16}\text{O}$ NCQE-like interaction.
Additional $\gamma$ rays may be produced within a time window up to a few tens of nanoseconds by the secondary interactions of hadrons. However, this timescale is too fast to be resolved by the SK detector.
On the other hand, the delayed signal is observed on much longer timescales, arising from $\gamma$ rays emitted when neutrons produced in the initial interaction or subsequent secondary interactions are captured by nuclei in the detector, predominantly gadolinium or hydrogen.
In this analysis, a 0.01\% concentration of Gd corresponds to a characteristic capture time of 115$\pm 1~\mu s$~\cite{Gd_2022}.
We note that, while the trigger defining the primary event is required to be consistent with the beam timing, there 
is no such restriction on the delayed signal.



\subsection{Prompt Event Search}

\begin{figure*}[htbp]
\centering
\includegraphics[width=\textwidth]{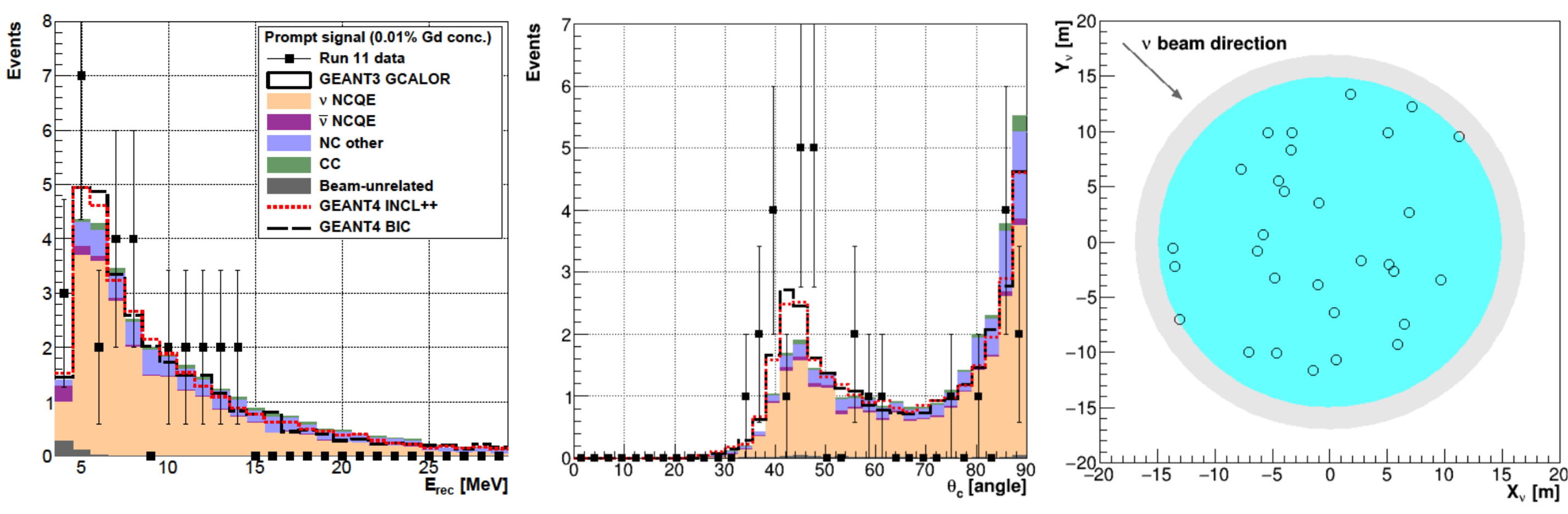}
\caption{Distributions of the reconstructed energy (\(\textit{E}_{\mathrm{rec}}\), left), Cherenkov angle (\(\theta_C\), middle), comparing data to MC prediction, and vertex position (right) for selected prompt events. The arrow in the right panel indicates the neutrino beam direction. The gray region marks the inner detector, and the blue region denotes the fiducial volume (top view).}
\label{fig:promptNCQE}
\end{figure*}
\begin{table*}[htbp]
\centering
\caption{Number of prompt event candidates surviving each selection step for the On-Beam and Off-Beam data, compared to various SI models predictions (\textsc{geant}\scalebox{0.8}{3} GCALOR and three SI models from \textsc{geant}\scalebox{0.8}{4}). The selection efficiency for each MC prediction is shown in parentheses. Off-Beam event counts are rescaled to the On-Beam time window length to allow direct comparison, while beam-unrelated backgrounds are not simulated. The final row, N$_{prompt}$, lists the number of prompt events passing all selections.}
\label{table:ncqedata}
\scalebox{1.1}{
\begin{tabular}{lccccccc}
\toprule
    & \multicolumn{2}{c}{Observed Data} && \multicolumn{4}{c}{Monte Carlo Simulation} \\
\cline{2-3} \cline{5-8}
{Selection}  & On-Beam &  Off-Beam && \textsc{geant}\scalebox{0.8}{3} GCALOR &  \textsc{geant}\scalebox{0.8}{4} BERT & \textsc{geant}\scalebox{0.8}{4} INCL++ & \textsc{geant}\scalebox{0.8}{4} BIC \\
\cline{1-3} \cline{5-8}
Timing    &  358   & 331.2  && - &-    &-    &-\\
Decay-e   &  356   & 330.9  && - &-    &-    &-\\
FV        &   92   &  67.4  && 35.9 (75.0\%) & 41.7(72.5\%) & 40.0(71.4\%) & 40.3(71.8\%)\\
$\textit{dwall}$   &   82   &  58.5  && 35.7 (74.6\%) & 41.5(72.1\%) & 39.8(70.9\%) & 40.0(71.4\%)\\
$\textit{effwall}$ &   60   &  34.7  && 35.3 (73.8\%) & 41.1(71.5\%) & 39.3(70.1\%) & 39.5(70.5\%)\\
$\textit{ovaQ}$    &   31   &   0.4  && 34.6 (73.7\%) & 39.5(68.7\%) & 37.5(66.9\%) & 37.8(67.4\%)\\
CC interaction &  30   & 0.4  && 31.5 (65.8\%) & 33.5(58.2\%) & 31.0(55.3\%) & 30.9(55.1\%)\\
\hline
\textit{N}$_{prompt}$ & 30 & 0.4  && 31.5  & 33.5 & 31.0 & 30.9\\
\hline
\end{tabular}
}
\end{table*}

The BONSAI reconstruction algorithm~\cite{Bonsai}, which is optimized for events with visible energies below approximately 50~MeV, is used to analyze the PMT hit pattern and timing of events within a 1.3~$\mu$s time window surrounding a trigger. 
Its vertex and direction reconstruction are based on a maximum likelihood method, assuming a single particle generated all detected light from a single vertex. 
A Cherenkov angle ($\theta_C$) for the event is obtained by calculating the cone angles between the event vertex and all hit PMT triplets within a 15~ns time trigger timing window, then defining $\theta_C$ as the peak of the resulting cone angle distribution.
The reconstructed energy for the event (\textit{E}$_{rec}$) is based on a relationship between energy and the calibrated PMT hits. 
Further details on the algorithm, its performance, and calibration are provided in Ref.~\cite{Bonsai}.


This study examines five event categories: neutrino NCQE interactions ($\nu$-NCQE), antineutrino NCQE interactions ($\bar{\nu}$-NCQE), other NC interactions (NC-other), CC interactions, and beam-unrelated backgrounds. 
The NC-other and CC categories incorporate both neutrino and antineutrino contributions. The NC-other category includes NC 1$\pi$ production, deep inelastic scattering, elastic scattering, and rare channels like $\ eta$- and K-meson production.
NEUT is used to simulate the first four interactions.
Beam-unrelated backgrounds, on the other hand, are estimated from the data set collected in the period from 500 to 5~\(\mu\text{s}\) before the beam arrives in SK.
This period is termed the Off-Beam time window.
Similarly, the On-Beam window is defined as the period from the beam spill arrival (0~ns) to 500~\(\mu\text{s}\) afterwards.
The event selection criteria below have been optimized to identify the prompt event from $\nu$-NCQE events and $\bar{\nu}$-NCQE events, while minimizing backgrounds from the other event categories.


The complete list of prompt event selection criteria are as follows:
\begin{enumerate}
    \item $\textit{E}_{rec}$ selection: prompt events must have a reconstructed total energy, $\textit{E}_{rec}$, in the range [3.49, 29.49]~MeV. The lower limit reflects the SK threshold, and the upper limit avoids Michel electron contamination.
    \item Timing selection: the reconstructed event time must be within $\pm$100~ns of the expected timing of a beam bunch. 
    \item Decay-e selection: events with more than 22 hits within a sliding 30~ns time window spanning 0.2 to 20~$\mu$s before the arrival time of the beam are rejected. This cut removes events created by the Michel electron from a preceeding muon or charged pion.
    \item FV selection: prompt events must fall within the fiducial volume; the reconstructed vertex is required to be more than 200~cm away from any ID wall.
    \item \textit{dwall}, \textit{effwall} and \textit{ovaQ} selection: to suppress backgrounds from radioactive impurities near the detector wall, additional cuts were applied in the [3.49, 5.99]~MeV reconstructed energy range. Cuts are based on three variables: the event’s distance to the ID wall (\textit{dwall}), the distance along the track to the ID wall (\textit{effwall}), and the reconstruction quality (\textit{OvaQ})~\cite{Bonsai}. Specifically, the following conditions must be met: 
    \begin{align}
    \textit{dwall} &> {\it p}^{\textit{dwall}}_{0} + {\it p}^{\textit{dwall}}_{1} \times {\it E}_{rec} , \\
    \textit{effwall} &> {\it p}^{\textit{effwall}}_{0} + {\it p}^{\textit{effwall}}_{1} \times {\it E}_{rec} , \\
    \textit{OvaQ} &> {\it p}^{\textit{OvaQ}}_{0} + {\it p}^{\textit{OvaQ}}_{1} \times {\it E}_{rec},
    \end{align}
     where the parameters have been optimized and are set as (${\it p}^{\textit{dwall}}_{0}$, ${\it p}^{\textit{dwall}}_{1}$) = (580.0~cm, $-$80.0~cm $\cdot$ MeV$^{-1}$), (${\it p}^{\textit{effwall}}_{0}$, ${\it p}^{\textit{effwall}}_{1}$) = (1941.5~cm, $-$314.0~cm $\cdot$ MeV$^{-1}$) and (${\it p}^{\textit{OvaQ}}_{0}$, ${\it p}^{\textit{OvaQ}}_{1}$) = (0.4125, 0.042~MeV$^{-1}$).
    \item CC interaction selection: in order to remove CC backgrounds, a cut on reconstructed energy and Cherenkov angle is applied. Events satisfying  \(\theta_C > \textit{a} \times \textit{E}_{rec} + \textit{b}\), are selected, where $\textit{a}$ = 1.67~$\mathrm{deg\cdot MeV^{-1}}$ and $\textit{b}$ = 15.0~$\mathrm{deg}$ have been optimized using MC.
\end{enumerate}
These selections are the same as those used in the past T2K NCQE analysis~\cite{NCQE_T2K_Run1-9}.
However, because the beam and detector conditions vary from run to run, the cut optimizations for criteria 5 and 6, based on SKDETSIM, have been updated in this study. We follow the procedure from the past T2K NCQE analysis to determine the cut parameters using a figure-of-merit (FOM), defined as:

\[
\text{FOM}=
\frac{\mathit{N}_{\mathrm{sig}}}{\sqrt{\mathit{N}_{\mathrm{sig}}+\mathit{N}_{\mathrm{bkg}}}},
\]
where \(\mathit{N}_{\mathrm{sig}}\) represents the number of MC predicted \(\nu\)-NCQE signal events, and \(\mathit{N}_{\mathrm{bkg}}\) is the total background. The total background consists of non-signal neutrino events from MC (e.g., NC-other and CC interactions) and beam-unrelated background events from Off-Beam data.

MC events are weighted using the delivered POT, the energy‑dependent neutrino flux, and the associated interaction cross section.
Table~\ref{table:ncqedata} summarizes the prompt event selection for both data and MC.
The final number of expected events from \textsc{geant}\scalebox{0.8}{3} GCALOR (SKDETSIM) is 31.9, which includes the contribution from beam-unrelated backgrounds (0.4), and 30 events are observed in the data. 
The effectiveness of the reduction of beam-unrelated backgrounds can be seen through each selection stage in the Off-Beam column.
After all selections, this background is reduced by about three orders of magnitude. 
Fig.~\ref{fig:promptNCQE} shows the $\textit{E}_{rec}$, $\theta_C$, and vertex distribution of the prompt events. 
The observed $\textit{E}_{rec}$ distribution agrees well with the predictions, and the event vertices are uniformly distributed in the detector, as is expected.
In the $\theta_C$ distribution, the data at $\sim$42$^\text{o}$ angles are above the MC expectation. 
A Kolmogorov-Smirnov test to the $\theta_C$ distribution yielded a p-value of 15\%, indicating acceptable compatibility of the data and MC.  
Appendix \ref{subsec:fea_with_n} provides a detailed look at the prompt events.

\subsection{Delayed Signal Search}

The delayed signal induced by a neutron capture is selected using an algorithm which consists of two stages: a pre-selection and a neural network (NN) classification. 
The pre-selection involves identifying PMT hit clusters that potentially represent neutron capture signals. 
Following this, the neural network classification is used to discriminate the neutron capture signal from accidental noise. 

\subsubsection{Pre-selection}
\begin{figure}[htbp]
\centering
\includegraphics[width=0.48\textwidth]{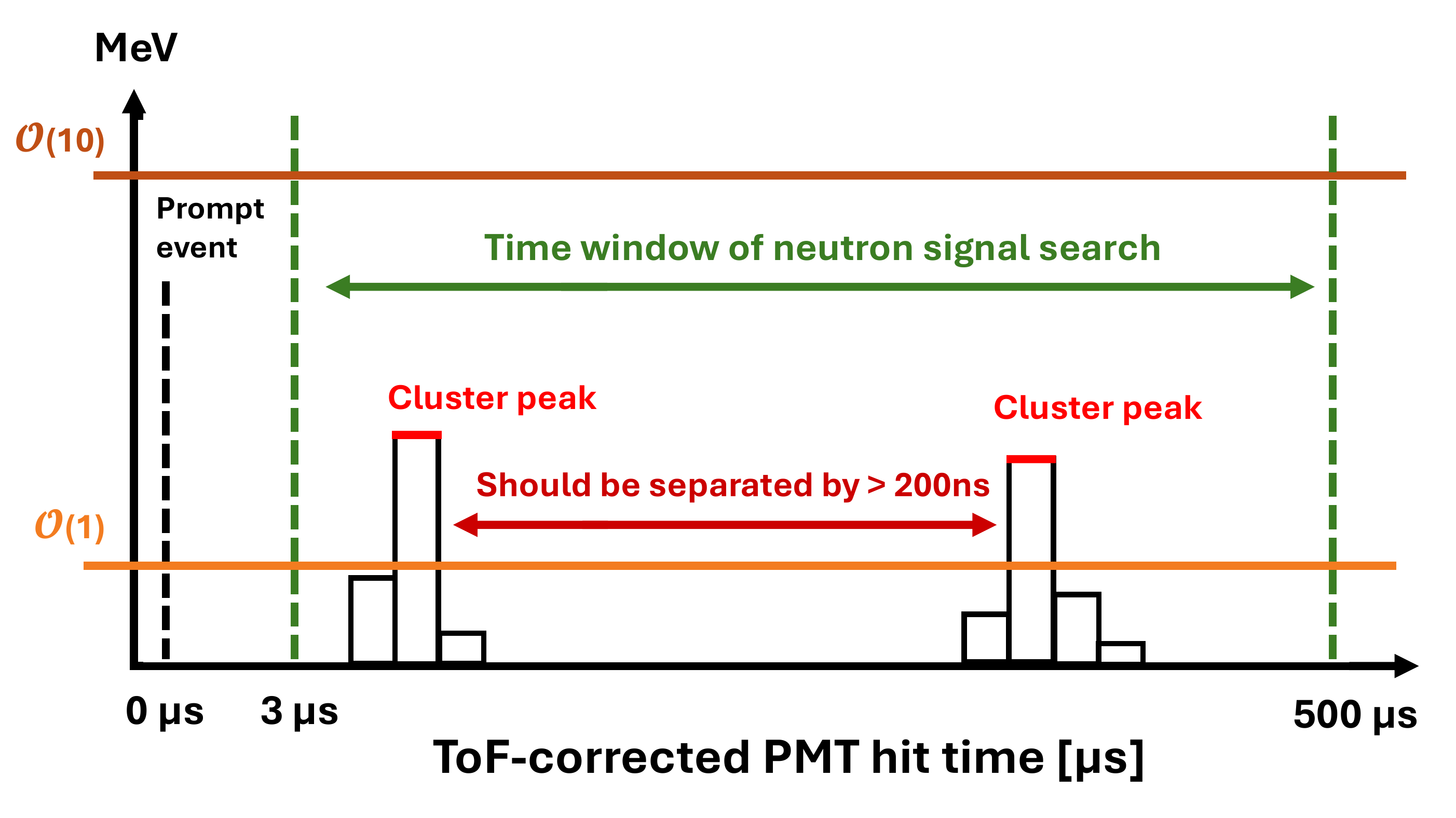}
\caption{Conceptual illustration of the neutron candidate pre-selection. PMT hits after each prompt event are grouped into ToF-corrected clusters within a [3, 500]~$\mu$s window. Clusters with reconstructed energy in the \(\mathcal{O}(1)\) $\sim$ \(\mathcal{O}(10)\)~MeV range are labeled as neutron candidates, with a minimum 200~ns gap to avoid double counting.}
\label{fig:n_preselect}
\end{figure}

Neutron candidates are searched for within [3, 500]~$\mu$s time window following a prompt event. 
Hit clusters are formed by finding a cluster of PMT hits with a time-of-flight (ToF) correction based on the prompt event vertex, which is reconstructed with BONSAI.
A 14~ns time window is employed for hit cluster searches, determined by the neutron capture time resolution and the optimization discussed in Ref.~\cite{Han_Thesis}.
Hit clusters with the number of PMT hits ranging between 7 and 400 PMT hits, which correspond to reconstructed energies between approximately $\mathcal{O}$(1) $\sim$ $\mathcal{O}$(10)~MeV, are classified as a neutron candidate.
This energy range includes the total energy deposition expected from neutron capture on both gadolinium and hydrogen with some margin. 
For events with one or more than one such candidate, a search for additional neutron candidates is performed. 
Each additional candidate is required to be more than 200~ns apart from a previous candidate to avoid double counting hits from previous clusters. 
Fig.~\ref{fig:n_preselect} illustrates the concept of this pre-selection.




\subsubsection{Neural network classification}

Each neutron candidate passing the pre-selection is then subject to a NN classification to reduce accidental backgrounds which typically arise from PMT dark noise. 
The 14 ToF‑corrected features were selected to capture key neutron candidate information and fall into five categories: hit‑cluster size (Hits), timing spread relative to the vertex (Timing), vertex quality and position (Vertex), topological correlations expected from Cherenkov cone (Topology), and PMT dark‑noise signatures (PMT Noise).

\begin{enumerate}[label=(\arabic*)]

    \item Hits: two variables related to the number of PMT hits, \texttt{NHits} and \texttt{NResHits}, are used.
    \texttt{NHits} denotes the number of PMT hits within a sliding time window with 14~ns width, with the window centered on the point of maximum hits. \texttt{NResHits} is the difference between the number of PMT hits taken in a larger window, [$-$100, +100] ns, and \texttt{NHits}.
    
    \item Timing: one variable related to the PMT timing, \texttt{TRMS}, is used. This parameter is the root-mean-square PMT hit time taken over all hits in the candidate's hit cluster. 
    
    \item Vertex: three distributions related to the candidate's vertex, \texttt{FitGoodness}, \texttt{DWall} and \texttt{DWallMeanDir} are used. 
    \texttt{FitGoodness} represents the quality of vertex reconstruction, based on the timing likelihood calculated by the reconstruction algorithm, assuming a PMT timing resolution of 5~ns. 
    \texttt{DWall} represents the shortest distance from the reconstructed vertex to any ID wall, and \texttt{DWallMeanDir} is the distance along the mean direction from the neutron candidate’s vertex to each hit PMT. 
    
    \item Topology: several parameters related to the spatial distribution of hits in the detector are used, $\text{\texttt{Beta}(\it{k})}$ with$\, \it{k} \in \{1, 2, 3, 4, 5\}$ and \texttt{OpeningAngleStdev}. 
     The $\text{\texttt{Beta}(\it{k})}$ parameters are defined as
     \begin{equation}
    \texttt{Beta}(\it{k}) \equiv \frac{2} { \mathrm{\texttt{NHits}} \left(\mathrm{\texttt{NHits}} - 1\right) } \sum_{i \neq j} P_{\it{k}}\left(\cos\theta_{\it{ij}}\right),
    \end{equation}
    where \( P_{\it{k}}(\cos \theta_{\it{ij}}) \) is the \( \it{k} \)-th degree Legendre polynomial, and \( \theta_{\it{ij}} \) represents the angle between a pair of PMT hits as viewed from the neutron candidate's reconstructed vertex.
     \texttt{OpeningAngleStdev} represents the standard deviation of the observed angle among triplets of hit PMT recorded within a 14~ns time window. 

    \item PMT Noise: two parameters related to the intrinsic noise characteristics of the PMT are used, \texttt{DarkLikelihood} and \texttt{BurstRatio}. 
    \texttt{DarkLikelihood} expresses the likelihood that the PMT hits were caused by dark noise and is calculated based on the measured dark rates of the individual hit PMT. 
    \texttt{BurstRatio} represents the ratio of PMT hits occurring in 10~$\mu$s prior to the neutron candidate relative to the total number of hits from the candidate. This parameter is used to separate neutron-induced hits from those created by scintillation light from particles emitted by radioactive impurities in the PMT glass.
    
\end{enumerate}

\begin{figure}[htbp]
\centering
\includegraphics[width=0.4\textwidth]{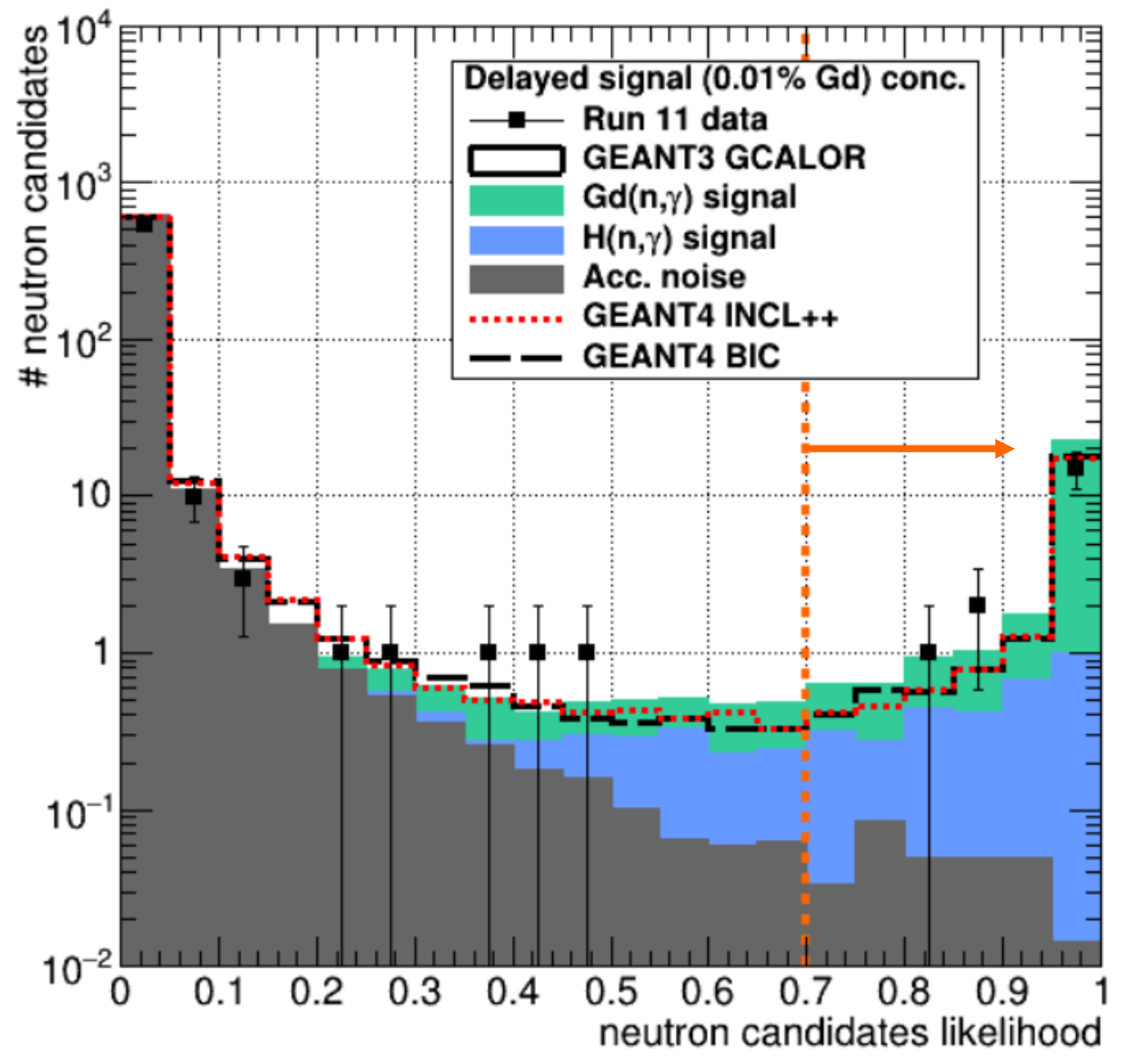}
\caption{The likelihood distribution of all neutron candidates from beam neutrino events, determined by the neural network. Neutron candidates with likelihood values above 0.7 are defined as ``delayed'' signals, matching the delayed signal counts in Table~\ref{tab:n_tn470data}, and yield approximately 99\% purity with about 1\% accidental noise.}
\label{fig:delayed_NNOut}
\end{figure}

\begin{figure*}[!htbp]
\centering
\includegraphics[width=\textwidth]{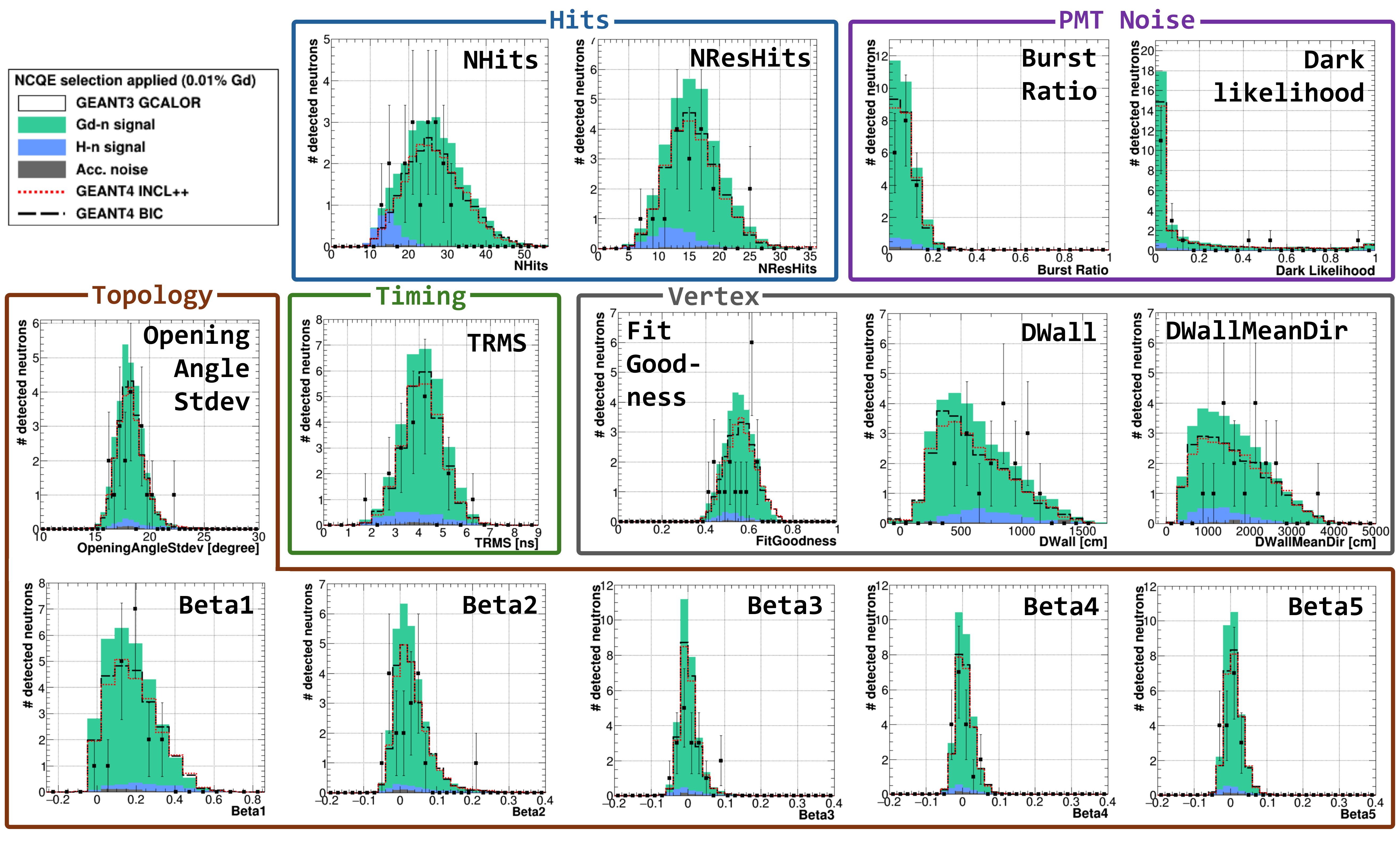}
\caption{Comparison of the 14 input features of the delayed signals for the data set and MC prediction in the neural network classification.} 
\label{fig:delayed_NN}
\end{figure*}

The feed-forward fully connected NN was trained and calibrated using a $^{241}$Am/Be neutron source and beam-unrelated background as described in Ref.~\cite{Han_Thesis}.  
Fig.~\ref{fig:delayed_NNOut} shows the likelihood output of neutron candidates from the beam data and MC. 
A neutron candidate is classified as a delayed signal if its likelihood output exceeds 0.7, achieving approximately 99\% neutron purity with the accidental noise rate per selected prompt event ($\eta_{Noise}$) around 1\%. 
The $\eta_{Noise}$ is calculated using the data set collected in the Off-Beam time window. 
Fig.~\ref{fig:delayed_NN} illustrates the 14 features of the delayed signals after the NN selection cut is applied. 
Overall, the neutron detection algorithm achieves a neutron detection efficiency ($\epsilon_{n}$) of 43.1\% from MC, as summarized in Table~\ref{tab:NTag_Eff}.  

\begin{table}[htbp]
\centering
\caption{Summary of the neutron detection algorithm for neutron detection efficiency, accidental noise rate, and purity.}
\label{tab:NTag_Eff}
\scalebox{1.0}{
\begin{tabular}{l|ccc}

& \multicolumn{2}{c}{{\it n} detection algo.} \\
\hline
& H({\it n},$\gamma$) & Gd({\it n},$\gamma$) \\
\hline
neutron detection efficiency   & 3.1\%   &  40.0\% \\
\hline
& \multicolumn{2}{c}{Overall} \\
\hline
neutron detection efficiency ($\epsilon_{n}$)  & \multicolumn{2}{c}{43.1\%} \\
Acc. noise rate ($\eta_{Noise}$)    & \multicolumn{2}{c}{1.3\%} \\
Purity                       & \multicolumn{2}{c}{98.7\%} \\
\end{tabular}
}
\end{table}



\begin{table*}[htbp]
\centering
\caption{Delayed signal numbers from the data set and MC are presented. Detector simulations with different SI models are provided, and ratios of each interaction channel to total signals are shown in parentheses. The data set is categorized into On-Beam and Off-Beam, where Off-Beam assesses accidental noise.}
\label{tab:n_tn470data}
\scalebox{1.1}{
\begin{tabular}{lcccrrrr}
\toprule
                 & \multicolumn{2}{c}{Observed Data} && \multicolumn{4}{c}{Monte Carlo simulation} \\
\cline{2-3} \cline{5-8}
{Channel}  & On-Beam & Off-Beam && \textsc{geant}\scalebox{0.8}{3} GCALOR &  \textsc{geant}\scalebox{0.8}{4} BERT & \textsc{geant}\scalebox{0.8}{4} INCL++ & \textsc{geant}\scalebox{0.8}{4} BIC \\
\cline{1-2} \cline{3-8}
\textrm{$\nu$-NCQE}         &  -  & -  && 19.3 (62.6\%) & 17.9 (53.8\%) & 12.9 (51.8\%) & 13.0 (51.6\%) \\
\textrm{$\bar{\nu}$-NCQE}   &  -  & -  &&  0.5 (1.6\%) &  0.5 (1.5\%)  & 0.4 (1.5\%) &  0.4 (1.5\%) \\
\textrm{NC-other}         &  -  & -  &&  9.5 (30.9\%) &  9.4 (28.2\%)  & 7.0 (28.2\%) &  7.1 (28.2\%) \\
\textrm{CC}               &  -  & -  &&  1.1 (3.7\%) &  4.6 (13.9\%)  & 3.9 (15.5\%) &  4.0 (15.8\%) \\
\textrm{Noise}            &  -  & 0.3 (1.7\%)  &&  0.4 (1.2\%) &  0.9 (2.6\%)  & 0.8 (3.0\%) &  0.8 (2.9\%) \\
\hline
\textit{N}$_{delayed}$              &  18  &  && 30.8  & 33.3   & 25.0 & 25.3 \\
\hline
\end{tabular}
}
\end{table*}

The results of the delayed signal search are summarized in Table~\ref{tab:n_tn470data}.
The \textit{N}$_{delayed}$ row represents the total number of delayed signals observed among the total number of prompt events, \textit{N}$_{prompt}$, shown in Table~\ref{table:ncqedata}.
The number of expected delayed signals is 30.8, while 18 are observed. 
The alternative SI models, INCL++ and BIC, predict a lower number than the BERT model, consistent with the NCQE study from atmospheric neutrinos in SK~\cite{NCQE_SKVI}.

\section{\label{sec:5_Result}Result}

This analysis measures the mean neutron capture multiplicity, \textit{M}, using the formula:

\begin{equation}
  \textit{M} = \frac{ (\textit{N}_{\text{delayed}} - \eta_{Noise} \times \textit{N}_{\text{prompt}}) / \epsilon_{\it n}}{\textit{N}_{\text{prompt}}}.
  \label{eq:Nmulti}
\end{equation}
The calculation begins with the number of delayed signals ($\textit{N}_{\text{delayed}}$). 
Since the observed $\textit{N}_{\text{delayed}}$ includes mis-tagged neutrons from accidental noise, its subtraction is required. 
For the data set, this is determined from the number of signals in the Off-Beam window, while for the MC, the number of mistagged neutrons is calculated by multiplying the accidental noise rate ($\eta_{Noise}$) by the number of prompt events ($\textit{N}_{\text{prompt}}$). 
After this subtraction the number of delayed signals is then corrected by the neutron detection efficiency ($\epsilon_{\it n}$).
Finally, the corrected number of delayed signals is divided by the number of prompt events ($\textit{N}_{\text{prompt}}$) to obtain \textit{M}.

The following sections present systematic uncertainties associated with the components of Eq.~(\ref{eq:Nmulti}). 


\subsection{Systematic Uncertainty of \texorpdfstring{$\epsilon_{\it n}$}{epsilon\_n}}
The neutron detection efficiency, $\epsilon_{\it n}$, is determined from simulations and is therefore subject to several modeling uncertainties,
which are summarized in Table~\ref{tab:Totalsystem}.
Among them, the detector response for the delayed signal is the dominant source, followed by those stemming from the prompt event selection and the nucleon SI model.
These three are discussed below, while detailed evaluations for the other error sources can be found in Appendix~\ref{Append:Systematic}.
\begin{table}[htbp]
\centering
\caption{Summary of the uncertainties contributing to the neutron detection efficiency ($\epsilon_{\it n}$). The dominant contribution arises from the detector response to delayed signals, while other sources introduce relatively smaller uncertainties.}
\label{tab:Totalsystem}
\scalebox{0.95}{
\begin{tabular}{l c}
\toprule
                & Uncertainty of $\epsilon_{\it n}$ \\
\midrule
Detector response to delayed signal  & $+$5.3\%/$-$8.1\% \\
Nucleon SI         & $+$0.0\%/$-$0.8\% \\
Prompt signal selection     & $\pm$0.5\% \\
\hline
MC statistics      & $\pm$0.1\% \\
$\nu$ beam flux    & $\pm$0.4\% \\
$\nu$ oscillation        & $\pm$0.01\% \\
$\nu$ interaction  & $\pm$0.4\% \\
$\pi$ capture on $^{16}$O & $+$0.0\%/$-$1.5\% \\
$\mu$ capture on $^{16}$O & $+$0.0\%/$-$0.5\% \\
PMT gain           & negligible \\
Water status       & negligible \\
\midrule
Total              & $+$5.4\%/$-$8.4\% \\
\bottomrule
\end{tabular}}
\end{table}


\subsubsection{Detector response to the delayed signal}
The uncertainty in the detector response to the delayed signal is assessed using a $^{241}$Am/Be neutron source. 
The most critical factors are the H({\it n}, $\gamma$) (Gd({\it n}, $\gamma$)) capture ratio, ${\it r}_H$(${\it r}_{Gd}$), the probability for thermal neutrons to be captured on hydrogen (gadolinium) and the accuracy of the gadolinium interaction modeling.

We find that the $^{241}$Am/Be MC simulation underestimates the ${\it r}_H$ compared with the calibration result.
By examining the PMT-hit distribution, we obtain a $\chi^2$-fitted ${\it r}_H$ of 56$\pm$3\%, which is consistent with the analytically predicted value of 56.2$\pm$1.5\% from ENDF/B-VII.1. In contrast, the $^{241}$Am/Be MC simulation predicts a lower value, 48\% for ${\it r}_H$~\cite{Han_Thesis}.
The underestimation of ${\it r}_H$ occurs because the \textsc{geant}\scalebox{0.8}{4} Neutron High-Precision model currently treats hydrogen as a free nucleus rather than accounting for water’s molecular mass. As a result, hydrogen is simulated to move 18 times faster, reducing the probability of neutron capture on hydrogen and thus lowering the ${\it r}_H$~\cite{Hino_neutron}. 
Since the re-calibration and further investigations of \textsc{geant}\scalebox{0.8}{3} are still on-going, this analysis uses the existing default settings.
After combining calibration and cross section uncertainties, we obtain ${\it r}_H$ with a central value of 48$^{+8.2}_{-1.5}$\%, and this uncertainty carries over to the ${\it r}_{Gd}$ because ${\it r}_{Gd} \approx 1-{\it r}_H$.
The $\epsilon_{\it n}$ is obtained by adding the hydrogen contribution, given by the product of ${\it r}_H$ and detection efficiency on hydrogen, to the gadolinium contribution, defined analogously. Propagating these correlated uncertainties yields an overall neutron detection efficiency of \(43.1^{+1.3}_{-6.1}\%\) in this study.
Table~\ref{tab:DetecRes} summarizes the uncertainty of $\epsilon_{\it n}$ related to the detector response to ${\it N}_{delayed}$.

\begin{table}[htbp]
\centering
\caption{Summary of systematic uncertainties of $\epsilon_{\it n}$ introduced by the detector response to the delayed signal. The table is derived from $^{241}$Am/Be neutron source calibration, encompassing neutron kinetic energy, detector, and Gd modeling~\cite{Han_Thesis}}.
\label{tab:DetecRes}
\scalebox{0.95}{
\begin{tabular}{c|l|c}
Source type      & Source name        &  \parbox{2.5cm}{Uncertainty of $\epsilon_{\it n}$} \\
\hline
AmBe neutron     & 1. neutron kinetic energy &  $\pm$0.3\% \\
modeling         &                           &        \\
\hline
Detector         & 2. Time evolution of      &  $\pm$0.6\% \\
modeling         &   \quad \ detector characteristics   & \\
                 & 3. PMT quantum efficiency & $\pm$0.4\% \\
\hline
Gd modeling      & 4. Gd/H-capture ratio     &  $+$1.3\%/$-$6.1\%\\
                 & 5. Gd-capture $\gamma$ model  & $\pm$5.1\% \\
\hline
\multicolumn{2}{c|}{Total} & $+$5.3\%/$-$8.1\%\\
\end{tabular} }
\end{table}

\subsubsection{Prompt event selection}
The systematic uncertainty on $\epsilon_{\it n}$ due to the prompt event selection includes contributions from cuts on $\textit{E}_{rec}$, $\theta_C$, $\textit{dwall}$, $\textit{effwall}$, and \textit{OvaQ}~\cite{Bonsai}, all of which carry uncertainties originating from the detector response and reconstruction. 
The uncertainties associated with $\textit{E}_{rec}$, $\textit{OvaQ}$, and $\theta_C$ are $5\%$, $1.5\%$, and $2$~degrees, respectively~\cite{SK_Calibration}. 
In addition, the uncertainties concerning $\textit{dwall}$ and $\textit{effwall}$ are derived from the vertex resolution.
The uncertainties in the vertex resolution along radial ($\textit{R}_{\nu}$) and vertical ($\textit{Z}_{\nu}$) directions are 10 cm and 5 cm, respectively~\cite{SK_Calibration}. 
Accordingly, the prompt event selection introduces an uncertainty of 0.5\% on $\epsilon_{\it n}$. 

\subsubsection{Nucleon SI} \label{subsubsec:nucleon_SI}
SI models differ in their adopted interaction cross sections and in their predictions of the outgoing nucleon kinematics, which may introduce a systematic uncertainty on $\epsilon_{\it n}$ and impact the prediction of neutron multiplicity.
In \textsc{geant}\scalebox{0.8}{3} GCALOR, the overall cross section in the energy range of interest is calculated based on nucleon-nucleon scattering cross sections~\cite{GCALOR}. 
An uncertainty of 30\% was assigned to this total cross section following Ref.~\cite{GCALORErr}, which compared the measured p-$^{12}$C scattering cross sections with several theoretical predictions. 
Notably, the prediction computed using the same nucleon SI cross sections adopted in this study agreed with the world experimental data on proton-carbon scattering to within 30\%, thus supporting the assigned error. 

In \textsc{geant}\scalebox{0.8}{4}, the uncertainty can be evaluated by performing the analysis using different SI models.
Since \textsc{geant}\scalebox{0.8}{3} GCALOR is based on the BERT model, we continue to use \textsc{geant}\scalebox{0.8}{4} BERT as our basis for the \textsc{geant}\scalebox{0.8}{4} analysis and estimate uncertainties by changing to the INCL++ and BIC models.

The uncertainty from the choice of SI model on $\epsilon_{\it n}$ is taken to be the largest change in detection efficiency among these models, $\Delta\epsilon_{SI}$ and estimated to be $^{+0.0\%}_{-0.8\%}$.
Table~\ref{tab:nucSI} summarizes the neutron detection efficiency, $\epsilon_{\it n}$, associated with the choice of SI model, along with the variations in the predicted number of delayed signals.

\begin{table}[htbp]
\centering
\caption{Change in detection efficiency, $\epsilon_{\it n}$, introduced by SI models with \textsc{geant3} GCALOR and \textsc{geant4} (BERT, INCL++, BIC). The neutron detection efficiency variation $\Delta\epsilon_{SI}$ is calculated by comparing the nominal to regenerated MC samples. The uncertainty in the predicted number of delayed signals is also summarized here.}
\label{tab:nucSI}
\scalebox{0.85}{
\begin{tabular}{lccccccc}
\toprule
    & \multicolumn{3}{c}{\textsc{geant}\scalebox{0.8}{3}} & &\multicolumn{3}{c}{\textsc{geant}\scalebox{0.8}{4}} \\ 
\cline{2-4} \cline{6-8} 
    & $-$30\% & GCALOR & $+$30\% && INCL++  & BERT  & BIC\\
\midrule
$\epsilon_{\it n}$     & 42.3\% & 43.1\% & 42.7\%  && 41.2\% & 41.5\% & 41.3\% \\
$\Delta\epsilon_{SI}$  & $-$0.8\% &  -  & $-$0.4\% && $-$0.3\% & - & $-$0.2\% \\
\cline{1-4} \cline{6-8} 
$\textit{N}_{delayed}$    & 28.4   & 30.8 & 31.6   &&  25.0   & 33.3 & 25.3 \\
$\Delta \textit{N}_{delayed}$  & $-$7.9\% &  -   & $+$2.5\% && $-$24.9\% & - & $-$24.0\% \\
\bottomrule
\end{tabular} }
\end{table}

\subsection{Mean Neutron Capture Multiplicity and Neutron Features}
The measured mean neutron capture multiplicity is:

{\small
\begin{align}
\textit{M}_{\text{data}} = 1.37 \pm 0.33 \, (\text{stat.})^{+0.17}_{-0.27}(\text{syst.}) 
\end{align}
}

\noindent The expectations extracted from MC with different SI models are as follows:
\begin{align}
\textit{M}_{\text{MC}}(\textsc{geant}\scalebox{0.8}{3}\;GCALOR) &= 2.24 \pm 0.01 \, (\text{stat.}) \\
\textit{M}_{\text{MC}}(\textsc{geant}\scalebox{0.8}{4}\;BERT) &= 2.28 \pm 0.01 \, (\text{stat.}) \\
\textit{M}_{\text{MC}}(\textsc{geant}\scalebox{0.8}{4}\;INCL++) &= 1.84 \pm 0.01 \, (\text{stat.}) \\
\textit{M}_{\text{MC}}(\textsc{geant}\scalebox{0.8}{4}\;BIC)  &= 1.87 \pm 0.01 \, (\text{stat.}) 
\end{align}

\begin{figure}[htbp]
\centering
\includegraphics[width=0.45\textwidth]{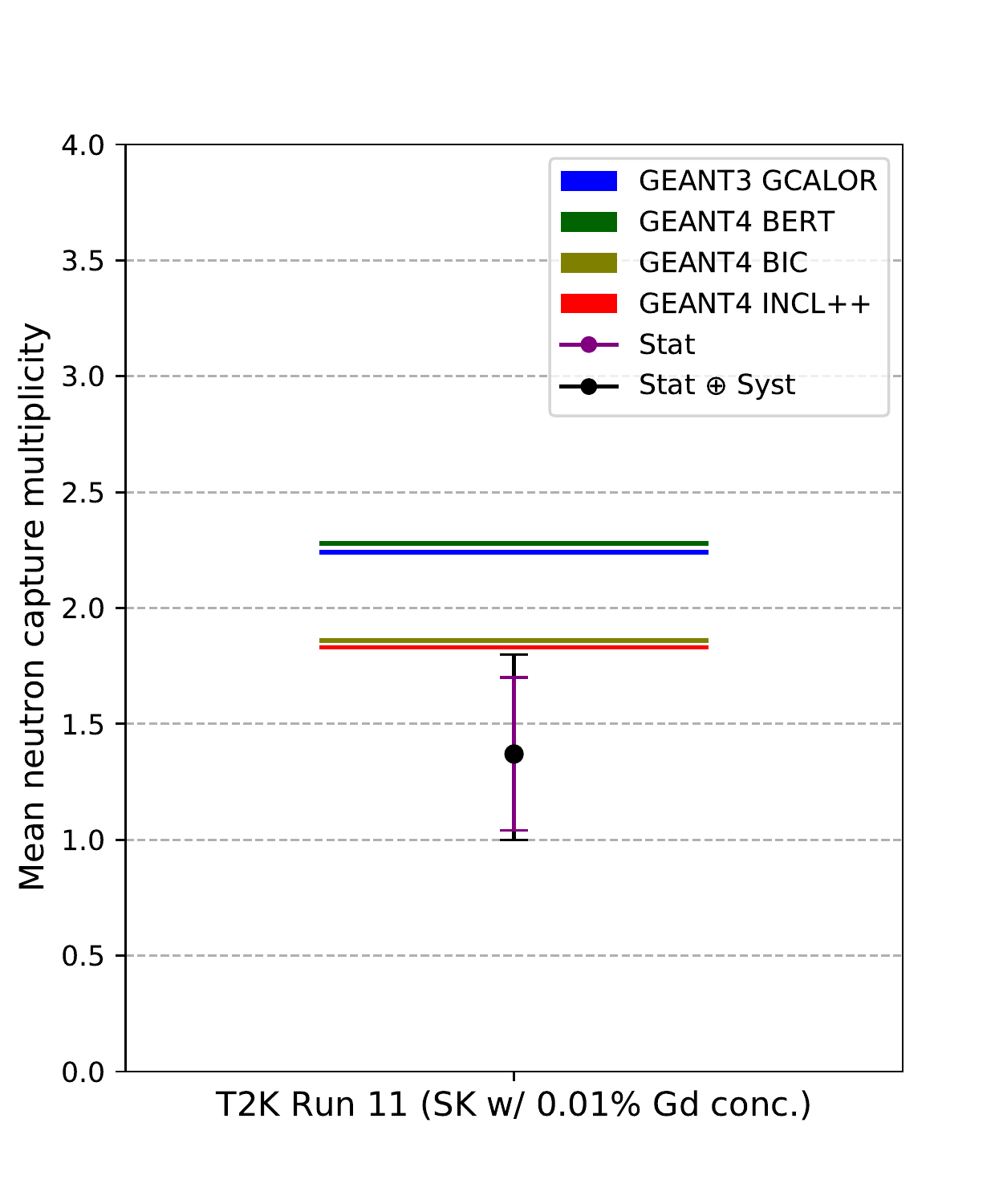}
\caption{Comparison of the measured mean neutron capture multiplicity to MC predictions using different SI models. The inner and outer error bars represent systematic and total uncertainties, respectively.}
\label{fig:n_multi_SI}
\end{figure}

\begin{figure}[htbp]
\centering
\includegraphics[width=0.42\textwidth]{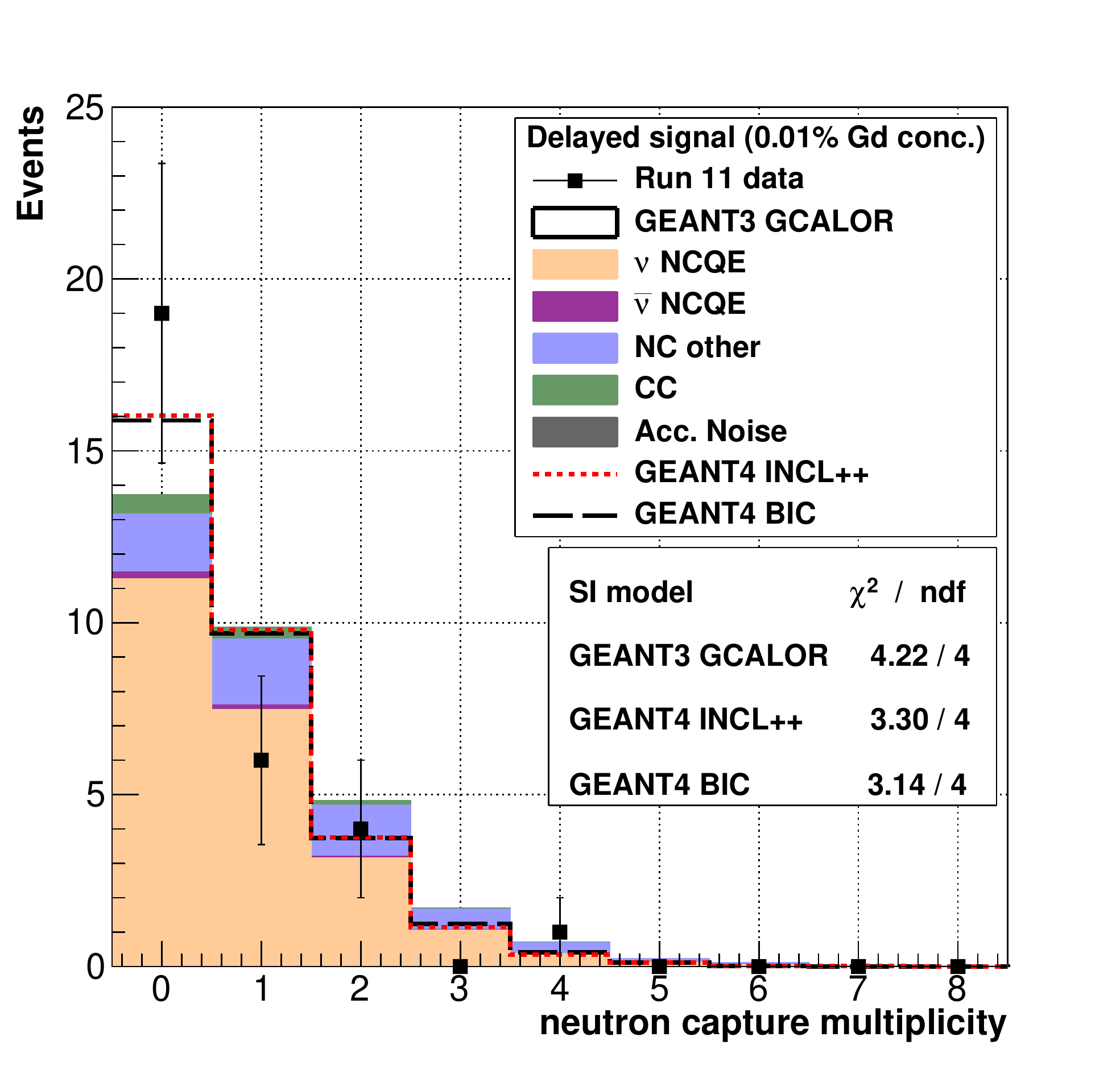}
\caption{Neutron capture multiplicity distribution for data compared to MC predictions with different SI models. }
\label{fig:n_multi}
\end{figure}

\noindent Fig.~\ref{fig:n_multi_SI} compares the mean neutron capture multiplicities obtained from this study with these expectations, while Fig.~\ref{fig:n_multi} shows the neutron capture multiplicity distribution. 
Note that all models overpredict the neutron multiplicity by at least $1\sigma$ compared to the data. 

Scaling the nominal SI cross section by 30\%  induces at most a 7.9\% change in the predicted number of delayed signals (c.f. Table~\ref{tab:nucSI}),
while changing the SI model in \textsc{geant}\scalebox{0.8}{4} leads to about a 25\% variation.
Uncertainties from other potential factors are also outlined in Table~\ref{tab:discussion}, but none has as significant an impact as the SI models in \textsc{geant}\scalebox{0.8}{4}. 
This result shows a slight preference for \textsc{geant}\scalebox{0.8}{4} INCL++ and BIC models, as they better reproduce the observed multiplicity and yield a lower $\chi^2$ value, as shown in Fig.~\ref{fig:n_multi}. 

\begin{figure*}[htbp]
\centering
\includegraphics[width=\textwidth]{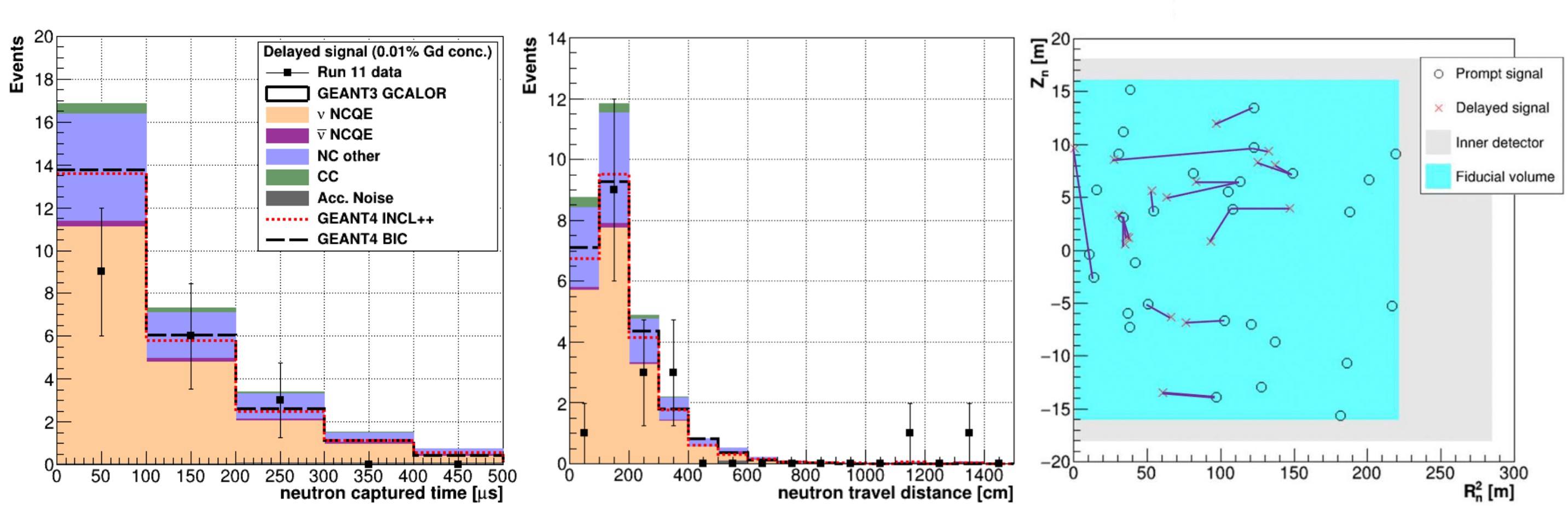}
\caption{Neutron capture time (left) and neutron travel distance (center) for data compared to MC predictions with different SI models. The right panel presents the two-dimensional distribution of NCQE-like events, with purple lines connecting each prompt event (circle) to its corresponding delayed signals (cross).}
\label{fig:n_feature}
\end{figure*}

The left plot of Fig.~\ref{fig:n_feature} presents the measured neutron capture time, while the center plot of Fig.~\ref{fig:n_feature} depicts the neutron travel distance, which is defined as the distance between the reconstructed prompt and delayed signal vertices. 
Although both figures demonstrate overall agreement between the data and MC, two delayed signals with travel distances more than 700~cm are observed though 0.5 signals are expected in this region. 

One of the delayed signals could be attributed to accidental noise, estimated at 0.4 signals in this study.
For signals beyond 700 cm, the Poisson probabilities of detecting 0, 1, or 2 neutrons are 60.7\%, 30.3\%, and 7.6\%; the corresponding noise probabilities are 67.0\%, 26.8\%, and 5.4\%. Combining these gives 9\% for one neutron plus one noise event.
However, we note that neutron travel distance is longer in data than in MC as indicated by the mismatch in the [0, 100] cm bin. 
Variation among SI models, which modify the neutron momentum spectra and consequently the predicted travel distances, does not fully explain this discrepancy.

Finally, the right plot of Fig.~\ref{fig:n_feature} shows the vertex distributions of selected prompt events and delayed signals with lines connecting the two when both exist.

\section{\label{sec:6_Discussion}Discussion}

Given the deficit of observed data relative to MC predictions for the number of delayed signals, we discuss the influence of modeling choices on those predictions. 
While other modelling aspects affect the prediction to a lesser extent than the adoption of SI model discussed above, we find that contributions for modeling of short-range nucleon correlations, NC 2p2h interactions, and the strange coupling constant induces a roughly 10\% variation. 
Other nuclear modelling, such as Pauli blocking and nuclear de-excitation processes, were found to have a minor impact. Table~\ref{tab:discussion} provides a summary of the delayed signal prediction variations from these studies.

\begin{table*}[htbp]
\centering
\caption{Summary of the delayed signal variations under different conditions beyond the SI models.}
\label{tab:discussion}
\scalebox{1.00}{
\begin{tabular}{l|c|l}
Factors                  & $\Delta {\it N}_{delayed}$  &  Reference setting \\
\hline
Nominal setting          & -                     &  NEUT w/ \textsc{geant}\scalebox{0.8}{3} GCALOR\\
Short-Range Correlations  & $+$9.4\%              &  NuWro~\cite{NuWro2109}       \\
NC 2p2h   & $+$12.0\%             &  NuWro w/ TEM model~\cite{2p2h_TEM} \\
$\textit{g}^{s}_A$                & $-$9.0\%              &  NuWro w/ $\textit{g}^{s}_A = -0.3$ \\
Pauli Blocking           & $-$0.5\%              &  NEUT w/ PB\\
Deexcitaion in $\nu$-$^{16}O$ NCQE & $+$0.9\%    &  NEUT w/ NucDeEx~\cite{Abe_NucDeEx} \\
\hline
\textsc{geant}\scalebox{0.8}{3} SI model                & $-$7.9\%/$+$2.5\%    &  NEUT w/ \textsc{geant}\scalebox{0.8}{3} GCALOR $\pm$30\% Xsec \\
\textsc{geant}\scalebox{0.8}{4} SI models                & $-$25.6\%    &  NEUT w/ \textsc{geant}\scalebox{0.8}{4} INCL++ \\
\end{tabular} }
\end{table*}

\subsection{Short-Range Correlations} 
\begin{figure}[htbp]
\centering
\includegraphics[width=0.45\textwidth]{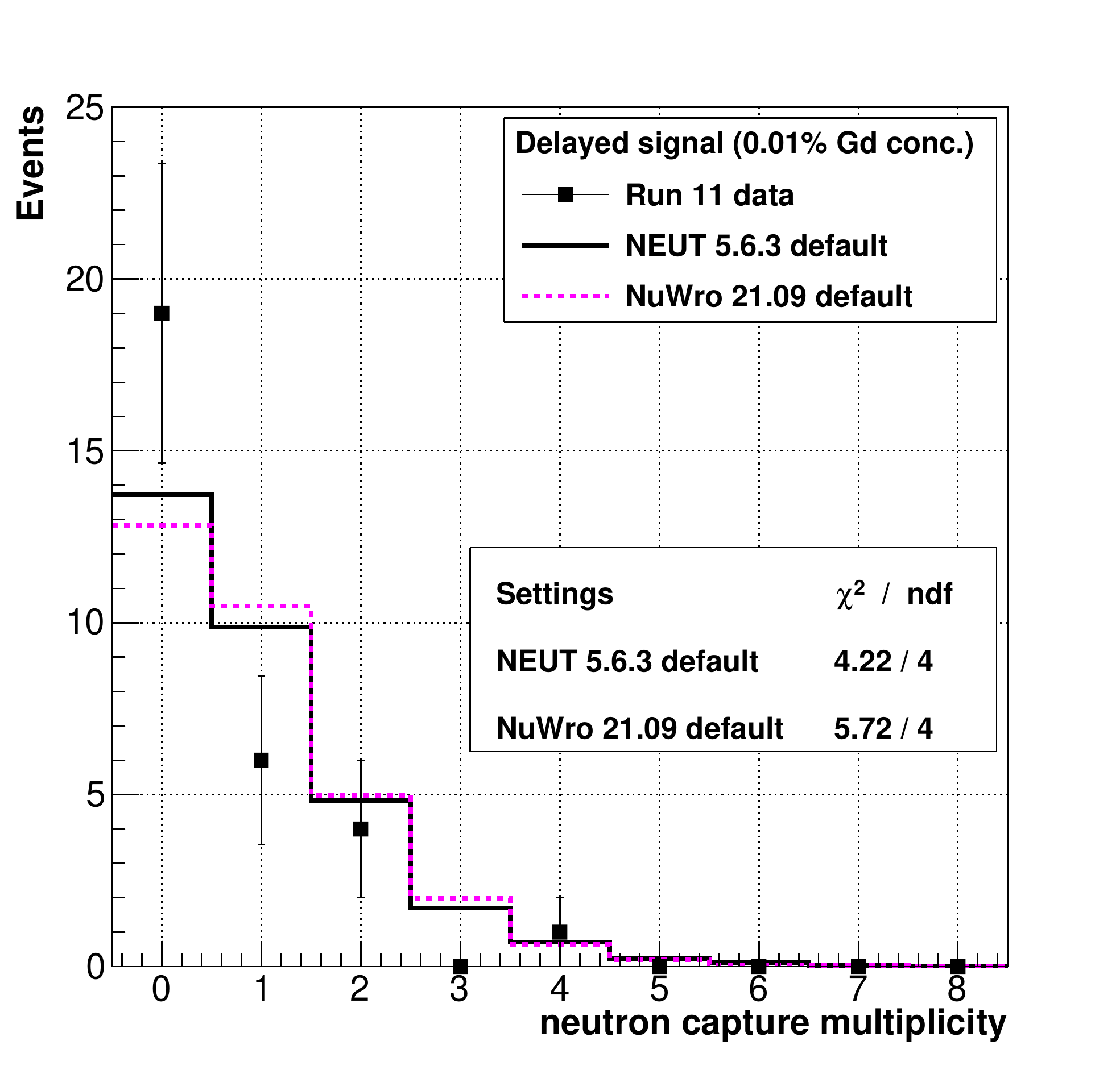}
\caption{The neutron capture multiplicity with NEUT default (black) and NuWro 21.09 default (red) is shown. NuWro predicts 9.4\% more neutrons than NEUT, which is a minor impact when compared to SI models.}
\label{fig:SRC}
\end{figure}

The spectral function by Benhar et al.~\cite{NCQE_Benhar} includes both mean-field and short-range correlations (SRC). 
SRC refers to the strong, localized interactions between pairs of nucleons within a nucleus.
Under the influence of SRC, a neutrino NCQE-like interaction can result in two outgoing nucleons from the primary interaction, both of which can go on to generate secondary neutrons via SI. 
NEUT and NuWro choose different SRC regions in the spectral function~\cite{SRC}, which may lead to different neutron predictions. 
Switching to events generated with NuWro (version 21.09), but propagated with the same SI model as the nominal analysis, yields 
the neutron multiplicity distribution shown in Fig.~\ref{fig:SRC}.
This change results in an overall increase of 9.4\% in the predicted number of neutrons.

\subsection{NC 2p2h}
\begin{figure}[htbp]
\centering
\includegraphics[width=0.45\textwidth]{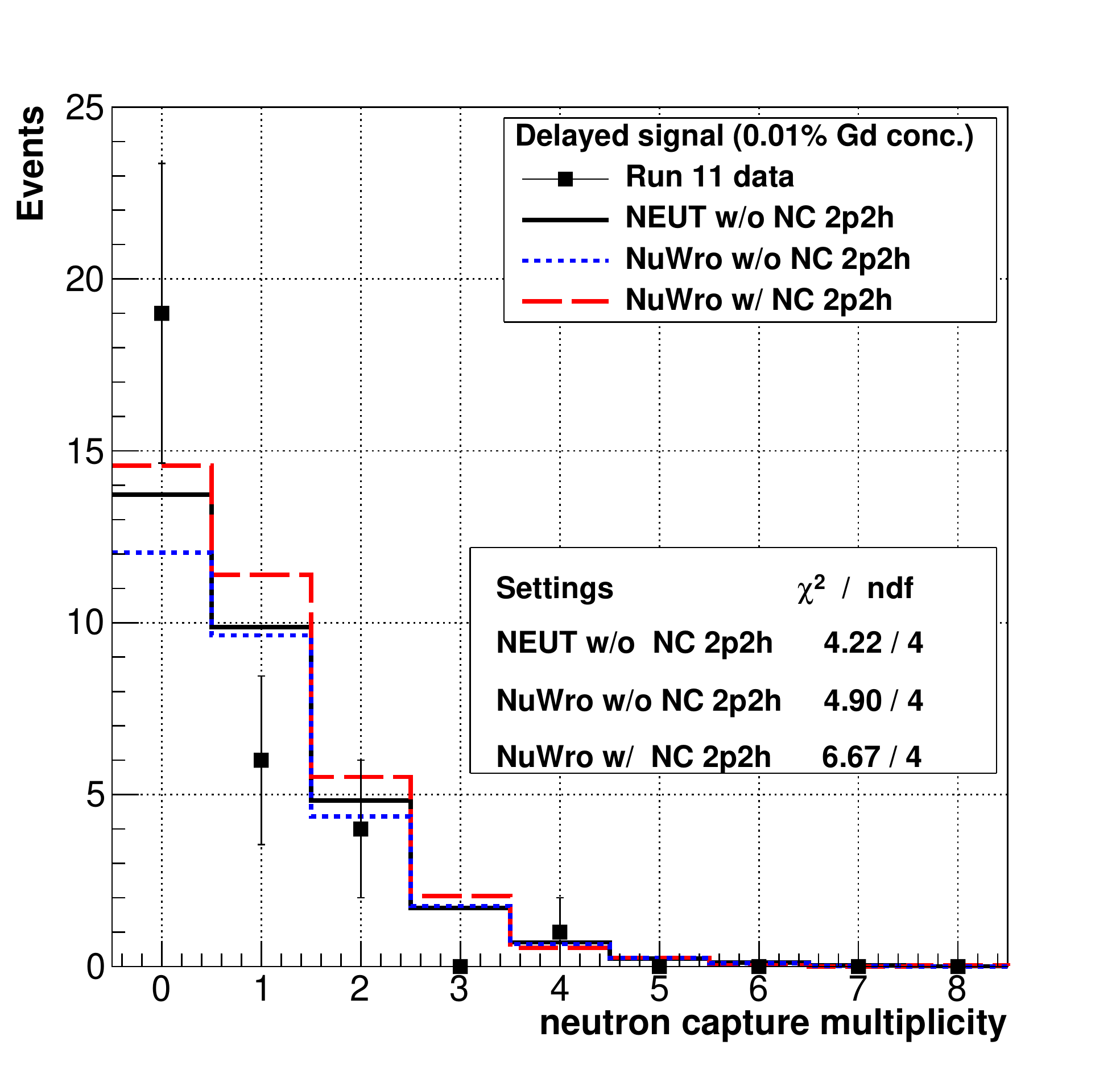}
\caption{The neutron capture multiplicity with the default NEUT prediction is represented by a black solid line. The NuWro predictions using the TEM model~\cite{2p2h_TEM}, with and without NC 2p2h interactions, are shown in red and blue dashed lines, respectively. When comparing the scenarios with and without the NC 2p2h interaction, there is an approximate 12\% increase in neutron prediction.}
\label{fig:nc2p2h}
\end{figure}

In the hypothesis of NC 2p2h interactions, both outgoing hadrons could lead to the production of neutrons in the final state, thereby changing the predictions used in this study. 
NEUT uses the Valencia 2p2h model by Nieves et al.~\cite{2p2h}, which does not include NC 2p2h interactions. 
In order to estimate the potential impact of adding the NC channel, we study the TEM model~\cite{2p2h_TEM} from NuWro.
The NuWro TEM model allows for the simulation of both CC and NC 2p2h interactions via the adjustment of vector magnetic form factors.
Importantly, the NC channel can be suppressed in the simulation. 
NuWro's predictions for the neutron capture multiplicity distribution with and without the NC 2p2h interaction are shown in Fig.~\ref{fig:nc2p2h} as dashed red and dotted blue lines, respectively. 
NEUT's prediction is depicted as a solid black line.
The number of delayed signals resulting from NC 2p2h interactions in the TEM constitutes an approximately 12\% increase.
For this comparison, the default value of the fraction of ${\it np}$ pairs, \(85^{+15}_{-20}\%\), was used. The fractions of ${\it pp}$ and ${\it nn}$ pairs, which share the same isospin, are roughly equivalent.

\subsection{Strange Axial Coupling Constant}
The neutron capture multiplicity could also be affected by the strange axial coupling constant ($\textit{g}_A^s$), which quantifies the contribution of strange quark-antiquark pairs to the nucleon's spin. 
NC processes involving the axial current could be influenced by \(\textit{g}_A^s\), and we focus on NCQE as it is the dominant contribution in this study.
A negative \(\textit{g}_A^s\) value enhances the NCQE cross section on protons but reduces it on neutrons~\cite{KamLAND}. 
Since NCQE interactions with protons seldom result in neutron emission, they generally lead to events with zero neutron capture multiplicity. 
As a result, a negative \(\textit{g}_A^s\) is expected to reduce the mean neutron capture multiplicity for NCQE interactions.

A negative \(\textit{g}_A^s\) means strange-quark spins anti-align with up/down quarks, lowering the nucleon’s total spin.
Investigations of \(\textit{g}_A^s\) using the NCQE interaction have been conducted in experiments such as BNL E734~\cite{BNL_E734}, MiniBooNE~\cite{MiniBooNE_2010} and KamLAND~\cite{KamLAND}, all of which favor negative values.
NEUT  version 5.6.3 uses  $\textit{g}_A^s = 0$ and does not support modifications to this parameter.
In contrast, NuWro allows for the direct input of \( \textit{g}_A^s \).
Here we use NuWro and scan \( \textit{g}_A^s \) from 0.0 to $-0.3$ to cover the majority of existing measurements from both neutrino and electron scattering experiments~\cite{EMC1,EMC2,HERMES,COMPASS}, with $-0.3$ serving as a conservative lower bound. 
Other generator settings are kept at their default values to evaluate the effect of \( \textit{g}_A^s \) on the neutron capture multiplicity.

\begin{figure}[htbp]
\centering
\includegraphics[width=0.45\textwidth]{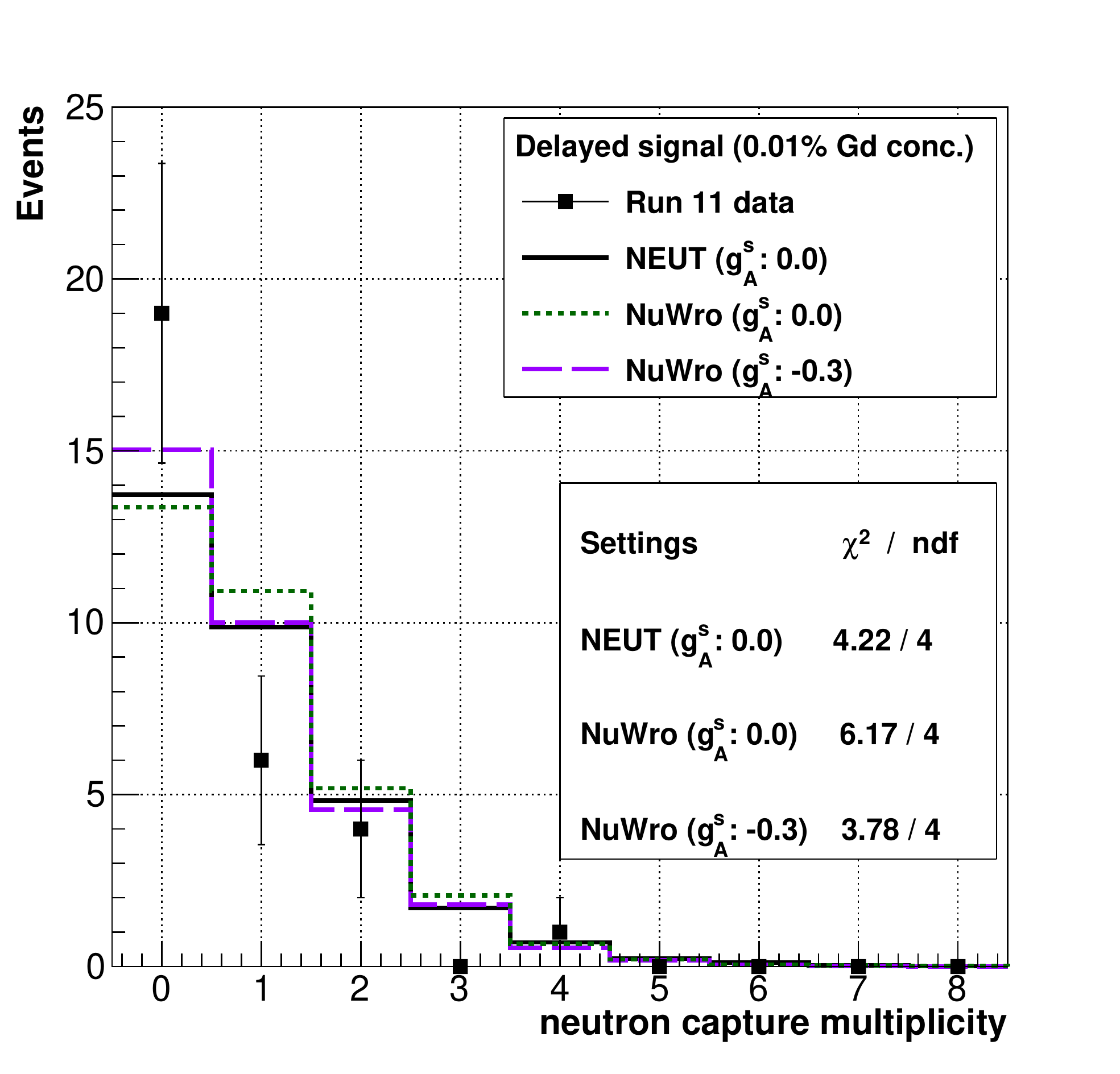}
\caption{The neutron capture multiplicity with the default $\textit{g}_A^s$ value in NEUT is represented by a black solid line. The NuWro predictions, with the default $\textit{g}_A^s$ value, 0.0, and the average values of existence \(\textit{g}_A^s\) measurement, $-$0.3, are shown in green dotted and purple dashed lines, respectively. When comparing the scenarios \(\textit{g}_A^s\) = $-$0.3 to \(\textit{g}_A^s\) = 0.0 within NuWro, there is an approximate 9\% decrease in the delayed signal prediction. 
}
\label{fig:gAs}
\end{figure}

Fig.~\ref{fig:gAs} shows the expected neutron capture multiplicity results for different \( \textit{g}_A^s \) values from NEUT and NuWro. 
The black line in Fig.~\ref{fig:gAs} represents the default \(\textit{g}_A^s\) value from NEUT, set at \(0.0\). 
The dotted green line indicates \(\textit{g}_A^s = 0.0\), the default value in NuWro. 
Finally, the dashed purple line represents \(\textit{g}_A^s = -0.3\). 
Setting \( \textit{g}_A^s \) to $-$0.3 reduces the predicted number of neutrons by 9\% compared to NuWro's default.

\subsection{Pauli Blocking}
The SF model~\cite{NCQE_Benhar, NEUT_Abe} adopted in NEUT 5.6.3 does not consider Pauli blocking. 
We therefore enforce it manually to estimate its impact on this analysis. 
Following the prescription in Ref.~\cite{PauliBlocking}, the cross section is set to zero in regions of phase space where, prior to undergoing FSI, the momentum of the outgoing primary nucleon falls below the Fermi momentum. 
Accordingly, this approach reduces the overall cross section and additionally results in a change in its shape at low momentum transfer. 
\begin{figure}[htbp]
\centering
\includegraphics[width=0.45\textwidth]{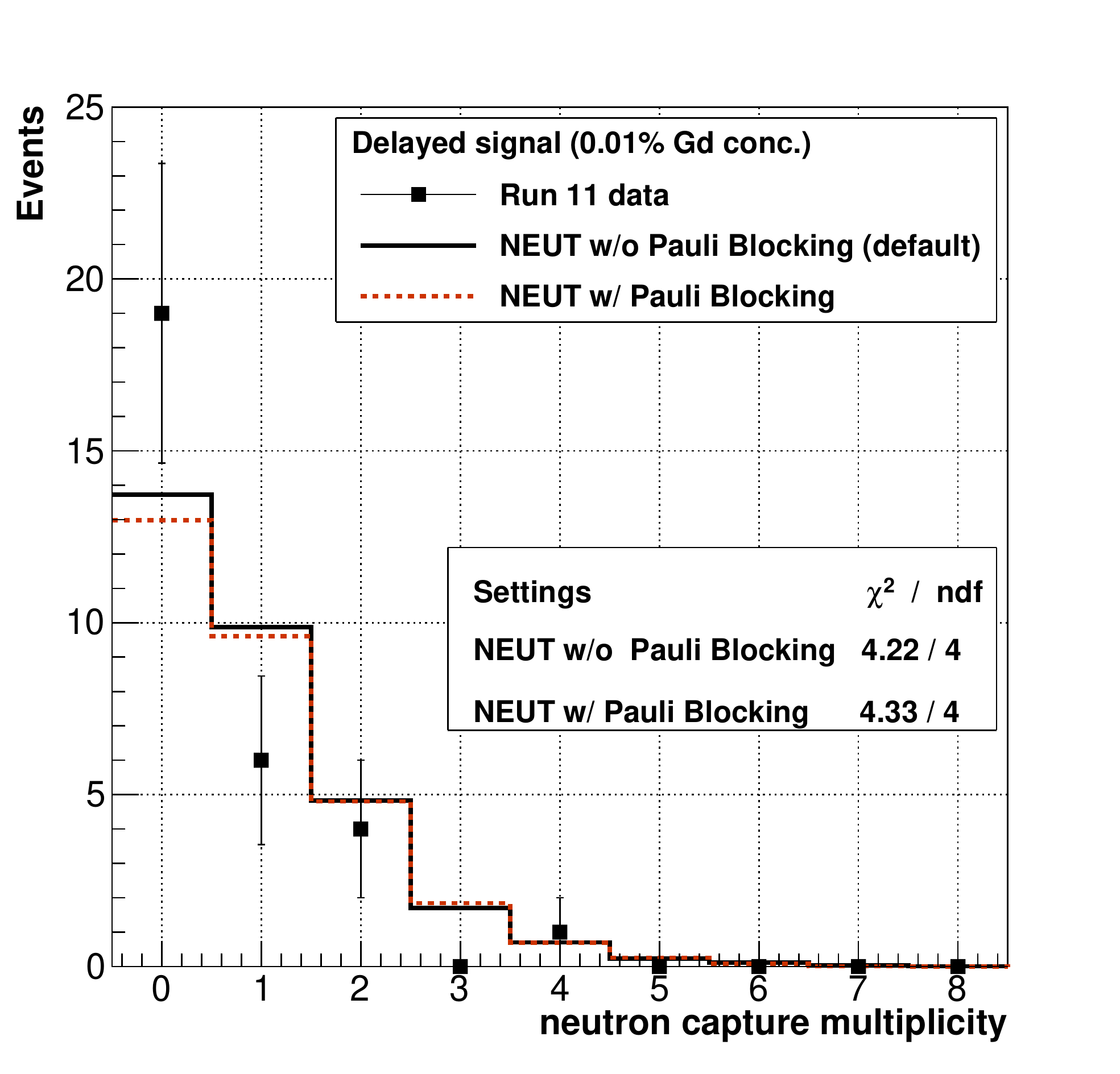}
\caption{The neutron capture multiplicity with and without Pauli blocking is shown as a black solid line and a brown dotted line, respectively. In this study, NEUT with Pauli blocking predicts approximately a 0.5\% reduction in the number of delayed signals compared to NEUT without Pauli blocking, which serves as the default.}
\label{fig:PB}
\end{figure}
Fig.~\ref{fig:PB} shows the neutron capture multiplicity prediction with (brown dotted line) and without (black) Pauli blocking enforced. 
Enabling Pauli blocking reduces the total neutron count by approximately 0.5\%, with predictions remaining nearly identical to the default. 
However, a decrease in the zero-neutron capture multiplicity is observed, suggesting a slight increase in the mean neutron capture multiplicity. 

\subsection{Nuclear De-excitation in Neutrino Interaction}
\begin{figure}[htbp]
\centering
\includegraphics[width=0.45\textwidth]{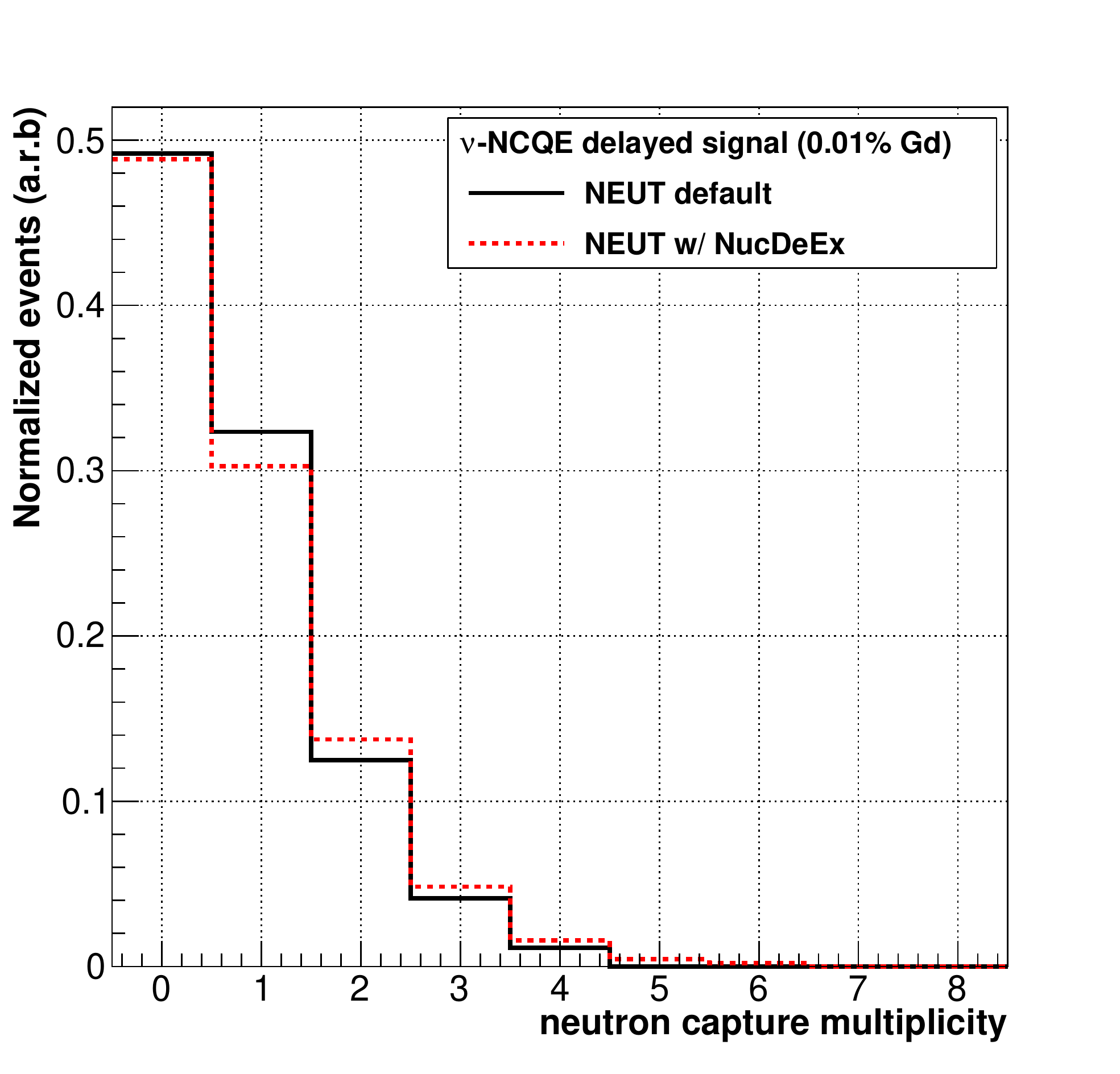}
\caption{The neutron capture multiplicity predictions with (red dotted line) and without (black solid line) NucDeEx are shown. The delayed signals prediction presented here are from $\nu$-NCQE interactions. NEUT with NucDeEx predicts a 0.9\% increase in delayed signals compared to the default de-excitation model in NEUT.}
\label{fig:NEUTNucDeEx_Nmulti}
\end{figure}

The choice of nuclear de-excitation model can also cause variations in the total number of predicted neutrons.
NEUT adopts a data-driven model~\cite{BRatio_Exp}, but we additionally consider the NucDeEx~\cite{Abe_NucDeEx, KamLAND_DeExDiscu} (version 1.0) model for an alternative estimate of the hadrons and photons produced during nuclear de-excitation. 
NucDeEx employs the nuclear reaction simulator TALYS version 1.96~\cite{TALYS} for the de-excitation calculation following a neutrino interaction with $^{16}\text{O}$.
TALYS has access to more recent, precise, complete data~\cite{Hauser-Feshbach} than the model used in NEUT. 
For this comparison, only $\nu$-NCQE interactions were considered in this study due to the limitations in NucDeEx.
Fig.~\ref{fig:NEUTNucDeEx_Nmulti} shows the neutron capture multiplicity with and without NucDeEx.
The simulation with NucDeEx resulted in a 5.1\% increase in the total number of generated neutrons compared to the case with the NEUT default de-excitation model without any selections. 
We noticed that the additional neutrons introduced by NucDeEx mostly occur in prompt events below 5 MeV, which is the main reason most of these neutrons are not retained. The relatively low energy of prompt events causes a significant loss of additional neutrons, both through the energy threshold applied in prompt event selection and through imprecise TOF corrections stemming from the relatively large vertex resolution during the pre-selection stage.
Overall, the full simulation with NucDeEx predicts a 0.9\% increase in the number of delayed signals relative to the nominal model.

\subsection{Insights from Other Neutrino-induced Neutron Measurements}
So far, only a few measurements of neutrons associated with neutrino interactions have been conducted. 
The T2K~\cite{Akutsu_thesis}, ANNIE~\cite{ANNRI_neutron}, MINER$\nu$A~\cite{MINERVA_neutron} and SNO~\cite{SNO_neutron} collaborations have all observed an overprediction of the neutron (capture) multiplicity with either beam or atmospheric neutrinos. 
These results collectively indicate that simulation settings with BERT-based models, which are used in these experiments, tend to overpredict the neutron (capture) multiplicity, while KamLAND~\cite{KamLAND}, whose simulation is based on the BIC model, does not see such an overprediction.
Given the wide range of energies spanned by these experiments, from MeV to GeV, neutron overprediction observed in both the CCQE and NCQE channels may hint at a connection to FSI modeling. 
However, as demonstrated above, the effect of SI modeling appears to dominate. 
SK~\cite{NCQE_SKVI} similarly tested several SI models, including BERT, INCL++, and BIC, using atmospheric neutrino NCQE events and reached a similar conclusion: BERT tends to overpredict, whereas INCL++ and BIC provide more accurate predictions for neutron capture multiplicity.
One of the main reasons is that BERT used the de-excitation model native to the Bertini code, while others use the G4PreCompound model~\cite{Sakai_thesis}. 
As described in Ref.~\cite{G4PreCompoundModel}, the G4PreCompound model relies on an exciton-based pre-equilibrium approach followed by semi-classical de-excitation mechanisms (e.g., Weisskopf-Ewing~\cite{Weisskopf} or GEM~\cite{GEM}) once the nucleus reaches equilibrium, which provides higher accuracy for lower-energy processes but requires more computational time.

\subsection{Prospects}

A precise understanding of NCQE interactions and their final state particles is essential for DSNB searches at SK, Hyper-Kamiokande~\cite{HyperK} and JUNO~\cite{JUNO}, as these detectors rely on neutron information to effectively reduce atmospheric NCQE backgrounds.
In addition, improving our understanding of NCQE events is essential for dark matter~\cite{DM_ACC2012, DM_ACC2017} and sterile neutrino search~\cite{SterileV_ACC2011, SterileV_ACC2017, SterileV_ACC2019} in accelerator neutrino experiments, which further motivates improvements to measurements such as this work. 

The statistical uncertainty on this measurement is expected to improve in the foreseeable future, as the anticipated POT proposed by T2K~\cite{T2K_POT} has not yet been reached. 

Furthermore, T2K data collected during the pure water~\cite{SKwaterNTag} and 0.03\% gadolinium~\cite{SK_2ndGdLoad} phases of SK could enhance the available statistics, providing roughly 10 times the neutron statistics used in this study.
Understanding secondary interactions then becomes essential to reduce uncertainties in this measurement.
Indeed, despite better data-MC compatibility in the current measurement when switching to the INCL++ and BIC models, there is still room for improvement.

While RCNP~\cite{RCNP_E487,RCNP_E525} and the ChipIr beamline at the Rutherford Appleton Laboratory~\cite{Chiplr} provide $\gamma$ measurements from SI, there remains a lack of neutron measurements from SI. 
An inverse-kinematics experiment using an oxygen beam at RIKEN-RIBF and the SAMURAI spectrometer~\cite{SAMURAI} can address this gap.
We anticipate that SI models will be better constrained in the near future by such measurements.
Dedicated upgrades to the \textsc{geant}\scalebox{0.8}{4}-based simulation may help reduce uncertainties on the neutron multiplicity and detection efficiency.
The flexibility of \textsc{geant}\scalebox{0.8}{4} allows users to customize the physics list based on the SI measurement mentioned above, offering the potential to enhance neutron kinematics predictions. A further dedicated calibration on the neutron capture, including the modification of \textsc{geant}\scalebox{0.8}{4} secondary interaction model, could further minimize uncertainties. 
While experiments like NINJA~\cite{NINJA} can help confirm CC 2p2h, direct measurements of NC 2p2h are an important missing piece. 
Electron scattering experiments, such as those at Jefferson Lab~\cite{JLab}, provide precise data on nucleon momentum distributions and could potentially improve short-range correlation modeling. 
These efforts will assist precise measurements of the strange axial coupling constant with neutrino sources, ultimately enhancing the accuracy of NCQE interactions. 

\section{\label{sec:7_Conclusion}Conclusion}

In this paper we presented the first measurement of neutron capture multiplicity resulting from $\nu$-$^{16}$O NCQE-like interactions, based on a data set corresponding to 1.76~$\times$~10$^{20}$ POT using the T2K neutrino beam.
The observed mean neutron capture multiplicity was \(1.37 \pm 0.33~(\text{stat.})^{+0.17}_{-0.27}~(\text{syst.})\).
The measurement deviates from the prediction made by NEUT combined with the Bertini-based SI model (\(2.24 \pm 0.01~(\text{stat.})\)). 
However, the predictions using other SI models, INCL++ (\(1.84 \pm 0.01~(\text{stat.})\)) and BIC (\(1.87 \pm 0.01~(\text{stat.})\)), are in closer agreement with the data.
We found that SI modeling dominates the uncertainty in the predicted numbers of neutrons, compared to modeling of processes, such as short-range correlations, NC 2p2h interactions, the strange axial coupling constant, Pauli blocking, and nuclear de-excitation following the neutrino interaction. 
Similar neutron overprediction has been observed in experiments using BERT-based SI models in both the CCQE~\cite{Akutsu_thesis, ANNRI_neutron, MINERVA_neutron, SNO_neutron} and NCQE~\cite{NCQE_SKVI} channels, that is not observed in measurements using BIC model~\cite{KamLAND}.
This work and these measurements highlight the need for a more precise understanding of SI processes and show that more data are needed to assess discrepancies and their origin.
Despite the current uncertainties, the neutron detection efficiency of NCQE-like events estimated in this study can be utilized in atmospheric neutrino studies, for example, by
accounting for the flux differences between T2K and atmospheric neutrinos. 
Further, the results of this study will serve as an essential validation tool for DSNB studies at water Cherenkov detectors, which aim to reduce atmospheric NCQE backgrounds using neutron information.

\section*{\label{sec:8_Acknowledgement}Acknowledgement}
The T2K collaboration would like to thank the J-PARC staff for superb accelerator performance. We thank the CERN NA61/SHINE Collaboration for providing valuable particle production data. We acknowledge the support of MEXT,   JSPS KAKENHI (JP16H06288, JP18K03682, JP18H03701, JP18H05537, JP19J01119, JP19J22440, JP19J22258, JP20H00162, JP20H00149, JP20J20304, JP24K17065) and bilateral programs (JPJSBP120204806, JPJSBP120209601),  Japan; NSERC, the NRC, and CFI, Canada; the CEA and CNRS/IN2P3, France; the Deutsche Forschungsgemeinschaft (DFG, German Research Foundation) 397763730, 517206441, Germany; the NKFIH (NKFIH 137812 and TKP2021-NKTA-64), Hungary; the INFN, Italy; the Ministry of Science and Higher Education (2023/WK/04) and the National Science Centre (UMO-2018/30/E/ST2/00441, UMO-2022/46/E/ST2/00336 and UMO-2021/43/D/ST2/01504), Poland;  the RSF (RSF 24-12-00271) and the Ministry of Science and Higher Education, Russia; MICINN  (PID2022-136297NB-I00 /AEI/10.13039/501100011033/ FEDER, UE, PID2021-124050NB-C31, PID2020-114687GB-I00,  PID2019-104676GB-C33), Government of Andalucia (FQM160, SOMM17/6105/UGR) and the University of Tokyo ICRR's Inter-University Research Program FY2023 Ref. J1, and ERDF and European Union NextGenerationEU funds (PRTR-C17.I1) and CERCA program, and University of Seville grant Ref. VIIPPIT-2023-V.4, and Secretariat for Universities and Research of the Ministry of Business and Knowledge of the Government of Catalonia and the European Social Fund (2022FI\_B 00336), Spain; the SNSF and SERI (200021\_185012, 200020\_188533, 20FL21\_186178I), Switzerland; the STFC and UKRI, UK; the DOE, USA; and NAFOSTED (103.99-2023.144,IZVSZ2.203433), Vietnam. We also thank CERN for the UA1/NOMAD magnet, DESY for the HERA-B magnet mover system, the BC DRI Group, Prairie DRI Group, ACENET, SciNet, and CalculQuebec consortia in the Digital Research Alliance of Canada, and GridPP in the United Kingdom, and the CNRS/IN2P3 Computing Center in France and NERSC (HEP-ERCAP0028625). In addition, the participation of individual researchers and institutions has been further supported by funds from the ERC (FP7), “la Caixa” Foundation  (ID 100010434, fellowship code LCF/BQ/IN17/11620050), the European Union’s Horizon 2020 Research and Innovation Programme under the Marie Sklodowska-Curie grant agreement numbers 713673 and 754496, and H2020 grant numbers  RISE-GA822070-JENNIFER2 2020 and RISE-GA872549-SK2HK; the JSPS, Japan; the Royal Society, UK; French ANR grant number ANR-19-CE31-0001 and ANR-21-CE31-0008; and  Sorbonne Université Emergences programmes; the SNF Eccellenza grant number PCEFP2\_203261;  the VAST-JSPS (No. QTJP01.02/20-22);  and the DOE Early Career programme, USA. For the purposes of open access, the authors have applied a Creative Commons Attribution license to any Author Accepted Manuscript version arising.

\appendix

\section{Uncertainties on Neutron Detection Efficiency}
\label{Append:Systematic}
\subsection{Statistical uncertainty}
Statistical uncertainty in the neutron detection efficiency stems from the finite size of the MC and is estimated as $\pm$0.1\%. 
Similarly, the accidental noise rate carries a $\pm$2.5\% statistical uncertainty due to the limited size of the Off-Beam data set.

\subsection{Neutrino beam flux}
Neutrino beam flux uncertainties  may cause variations in the types of neutrino interactions occuring in the detector, which subsequently 
affects our neutron detection efficiency. 
These uncertainties are evaluated for each flavor, energy, and horn polarity~\cite{T2K_NA61_2019}. 
Hadron production and interaction modeling account for $\sim$8\% around the flux peak, which represents the largest source. 
This study estimates the neutrino beam flux uncertainty in the same manner as the previous T2K analysis~\cite{NCQE_T2K_Run1-9}, 
resulting in a $\pm$0.4\% uncertainty on the neutron detection efficiency.

\subsection{Neutrino oscillation}
Uncertainties in oscillation parameters can affect the sample composition and, in turn, the neutron detection efficiency. 
The oscillation parameters and their associated uncertainties are sourced from Ref.~\cite{T2K_oscilla_para}.
Accounting for these errors induces only a $\pm$0.01\% uncertainty on the neutron detection efficiency since only CC events experience 
an observable oscillation effect.

\subsection{Neutrino interaction model}
The parameters and errors that describe the neutrino interaction cross sections and their values are taken from Ref.~\cite{Hadrons_FSI}. 
The value of the axial-vector mass used to generate quasielastic interactions with its 1\(\sigma\) error is \(\textit{M}_A^{\text{QE}} = 1.21 \pm 0.18\)~GeV/\(c^2\), while the Fermi momentum in oxygen is \(209 \pm 31\)~MeV/\(c\). 
Parameters describing contributions from 2p2h interactions, resonant pion production, and deep inelastic scattering follow the assignments in the previous analysis~\cite{NCQE_T2K_Run1-9}.
These parameters' uncertainties induce a $\pm$0.1\% uncertainty on the neutron detection efficiency. 

Additionally, nucleon FSI uncertainties could cause variations in the number and energy spectrum of neutrons coming from the $\nu$-$^{16}$O interaction, thereby affecting the neutron detection efficiency. 
We vary the scattering and particle production probability of the \textit{hA} model~\cite{GENIE_2015hA} within GENIE~\cite{GENIE} to assess the possible impact.
Assigning  a 20\% uncertainty to the total rescattering probability, 50\% to the charge exchange probability, 30\% to the elastic scattering probability, 
40\% to the inelastic scattering probability, 20\% to the absorption probability, and 20\% to the pion production probability results in an uncertainty of $\pm$0.4\% on the neutron detection efficiency. 

\subsection{\texorpdfstring{$\mu$ and $\pi$ capture on $^{16}$O}{Pion capture on 16O}}
A small portion of CC interactions may generate muons that can be captured by oxygen atoms and consequently produce additional neutrons.
Likewise, some neutral current interactions can create pions, whose capture may also produce neutrons. 
We evaluate the uncertainty from these processes by switching from the intra-nuclear cascade model to CHIPS~\cite{CHIPS} within SKDETSIM. 
The resulting uncertainties on the neutron detection efficiency are:  $^{+0.0\%}_{-1.5\%}$ for $\pi^{-}$ capture on $^{16}O$ and $^{+0.0\%}_{-0.5\%}$ for $\mu^{-}$ capture. 

\section{Application to Astrophysical Anti-neutrino Searches} \label{sec:DSNB_application}

The T2K NCQE sample presented in this work is a unique, high-purity sample of NCQE-like events with all of the same kinematic features as the atmospheric NCQE 
background to DSNB searches in water Cherenkov detectors. 
It therefore represents a valuable tool for assessing systematic uncertainties in those searches. 
In the following we discuss features of T2K that will support DSNB searches conducted at SK.

\subsection{Neutron detection algorithm used in the SK DSNB search}
Various neutron detection algorithms are available at SK. 
Focusing on the neutron detection algorithm used in the DSNB search at SK (termed $n_{DSNB}$ below)~\cite{Nu2024_DSNB} we evaluate its performance 
with the neutron detection algorithm employed in this study (termed $n_{T2K}$). 

The two algorithms diverge in how they handle the neutron candidate vertex, which is a key factor in the time-of-flight (ToF) calculation for neutron candidate features. 
As most neutrons stay close to the neutrino interaction vertex, $n_{DSNB}$ adopts the prompt event vertex for the ToF correction.
This approach facilitates the search for hydrogen-capture neutrons, H(${\it n},\gamma$), whose vertex is challenging to reconstruct directly due to the low number of PMT hits. 
However, as the neutron travels farther, the prompt event vertex becomes less accurate for ToF corrections, which results in the neutron detection efficiency depending upon the neutron kinematics.

\begin{figure}[htbp]
\centering
\includegraphics[width=0.4\textwidth]{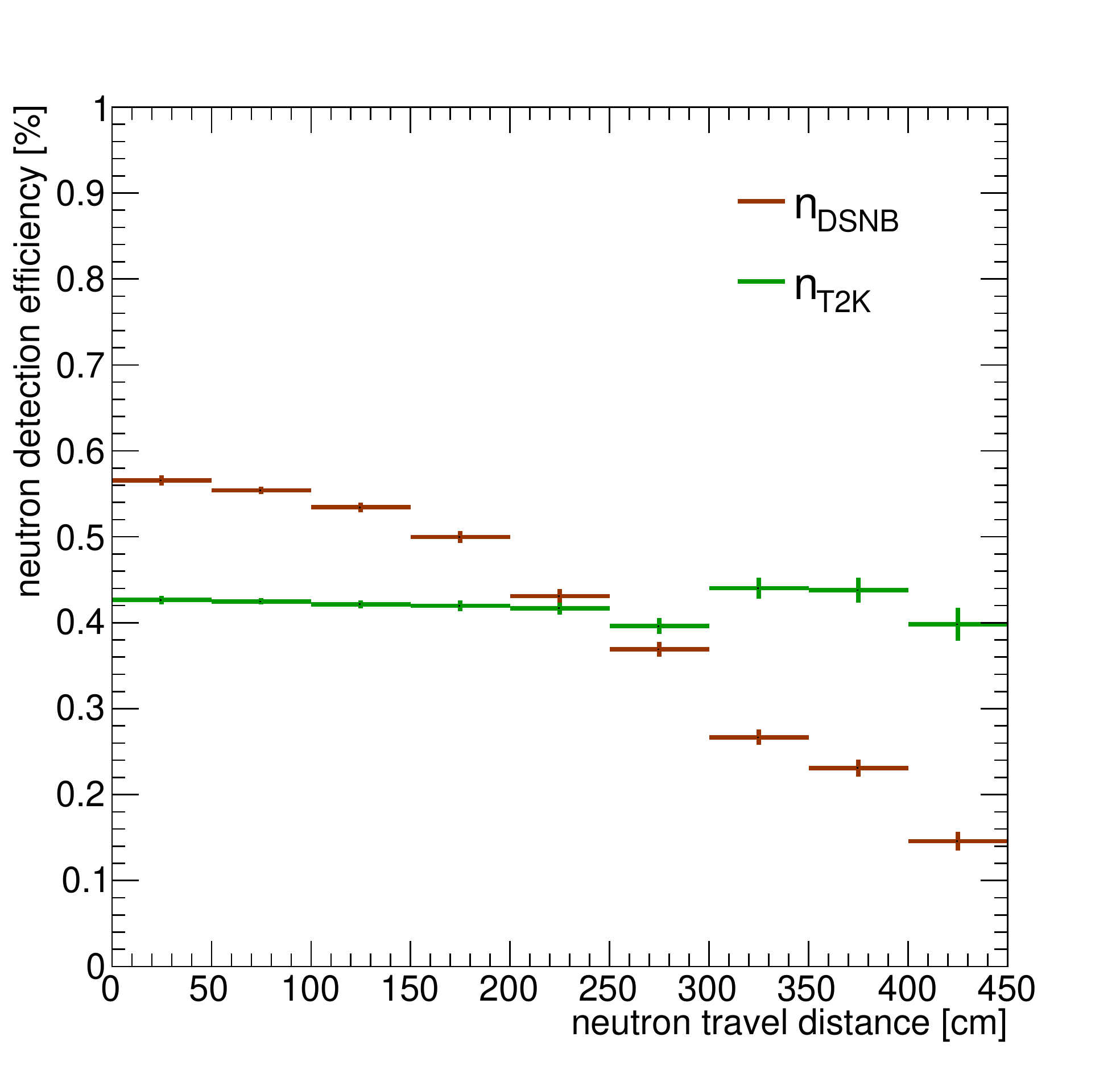}
\caption{Neutron detection efficiency versus neutron travel distance for $n_{T2K}$ and $n_{DSNB}$, with statistical error bars. In $n_{DSNB}$, the prompt event vertex is used for ToF corrections, making the efficiency more dependent on neutron kinematics. By contrast, $n_{T2K}$ employs the neutron candidate's own vertex for ToF corrections and thus remains relatively independent of the neutron travel distance.}
\label{fig:DisNeuEff}
\end{figure}

On the other hand,  $n_{T2K}$ adopts the reconstructed neutron candidate vertex as the vertex for the ToF correction, making it independent of the neutron kinematics.
Fig.~\ref{fig:DisNeuEff} shows the neutron detection efficiency as a function of the neutron travel distance for both algorithms and confirms the kinematic independence of $n_{T2K}$. 
Since SI modeling can affect predicted neutron kinematics, maintaining kinematic independence helps constrain systematic uncertainties associated with SI models in this study. 

Table~\ref{tab:NTag_Efficiency} compares the performance of $n_{T2K}$ and $n_{DSNB}$. 
Compared to $n_{T2K}$, $n_{DSNB}$ achieves a 7.8\% higher neutron detection efficiency, though it incurs slightly larger systematic uncertainties due to its dependence on neutron kinematics. 
Both algorithms perform similarly in terms of purity and \(\eta_{Noise}\). 
In order to remain consistent with the DSNB search at SK, the following sections employ $n_{DSNB}$ for further analysis.
\begin{table}[htbp]
\centering
\caption{Summary of $n_{T2K}$ and $n_{DSNB}$ for neutron detection efficiency, SI uncertainty, accidental noise rate, and purity. $n_{DSNB}$ offers higher detection efficiency, particularly for H({\it n},$\gamma$), while the $n_{DSNB}$ is less dependent on neutron kinematics and, thus, better at controlling systematic uncertainties.}
\label{tab:NTag_Efficiency}
\scalebox{0.85}{
\begin{tabular}{l||ccc|cc}

& \multicolumn{2}{c}{$n_{T2K}$}  & &\multicolumn{2}{c}{$n_{DSNB}$} \\
\hline
& H({\it n},$\gamma$) & Gd({\it n},$\gamma$) & &H({\it n},$\gamma$) & Gd({\it n},$\gamma$) \\
\hline
neutron detection efficiency   & 3.1\%   &  40.0\% & & 8.4\%  & 42.5\%  \\ 
\hline
& \multicolumn{2}{c}{Overall}  & &\multicolumn{2}{c}{Overall} \\
\hline
neutron detection efficiency ($\epsilon_{n}$)  & \multicolumn{2}{c}{43.1\%}  & & \multicolumn{2}{c}{50.9\%} \\
SI model syst. on $\epsilon_{n}$ & \multicolumn{2}{c}{$+0.0\% / -0.8\%$}  & & \multicolumn{2}{c} {$+1.7\% / -3.4\%$} \\
Acc. noise rate ($\eta_{Noise}$)    & \multicolumn{2}{c}{1.28\%}  & & \multicolumn{2}{c}{1.47\%} \\
Purity                       & \multicolumn{2}{c}{98.7\%}  & & \multicolumn{2}{c}{98.3\%} \\
\end{tabular}
}
\end{table}

\subsection{Prompt events features with neutron detection}
\label{subsec:fea_with_n}
The reconstructed energy and Cherenkov angle distributions of T2K NCQE-like events have been studied in detail previously~\cite {NCQE_T2K_Run1-9}.
However, this is the first time that neutron detection has been applied to T2K NCQE-like events.
Since the DSNB signal includes a neutron in the final state, the present sample is unique for examining the kinematic features of NCQE-like events accompanied by neutrons. 

\begin{figure*}[t]
\centering
\includegraphics[width=0.65\textwidth]{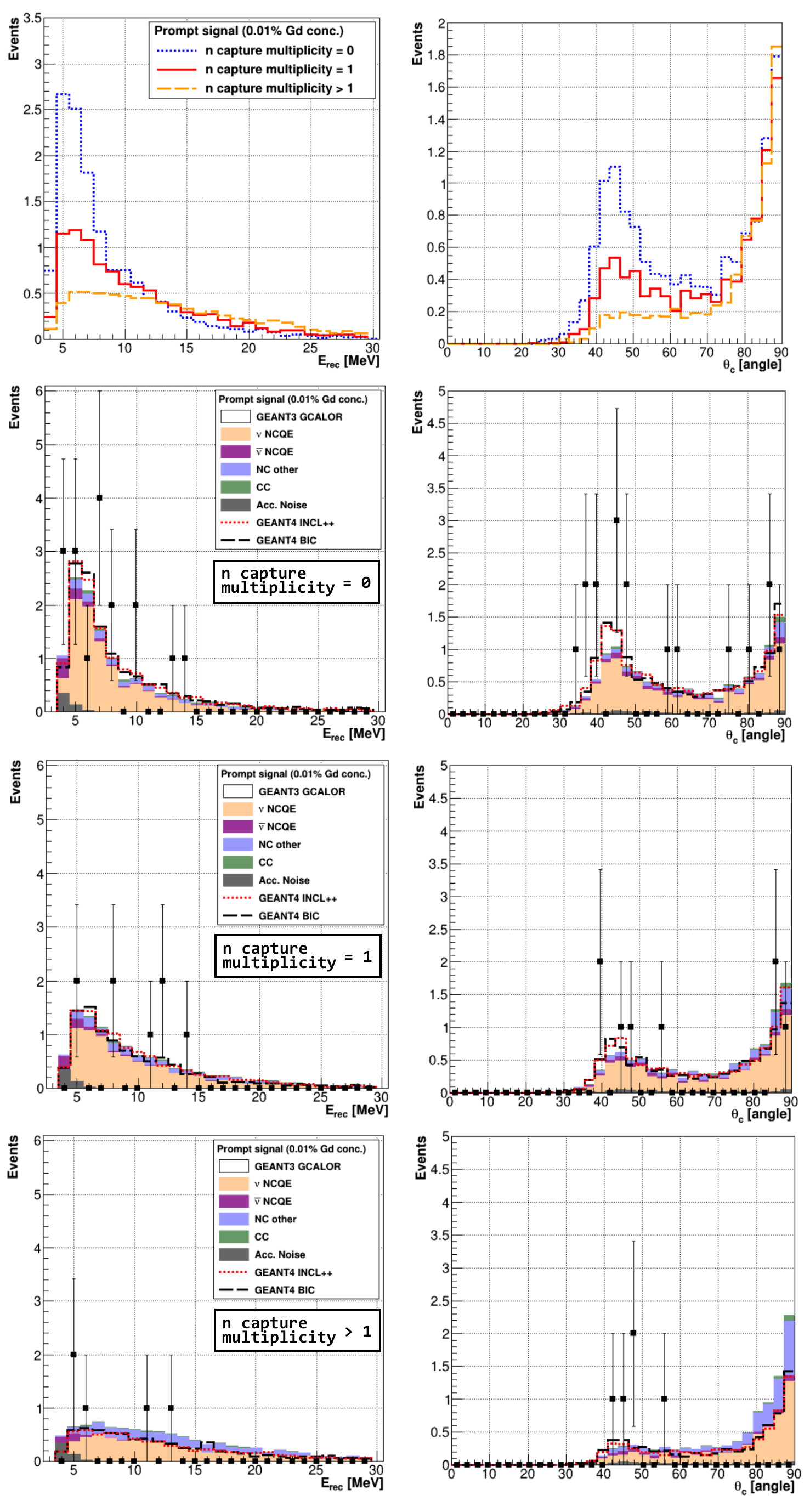}
\caption{Reconstructed energy and Cherenkov angle distributions for NCQE-like prompt events are shown, categorized by neutron capture multiplicity (zero: dotted blue, one: solid red, more than one: dashed orange) in the top row. The data set is compared with MC predictions incorporating different SI models, with neutron capture multiplicity information displayed in the second row (zero), the third row (one), and the bottom row (more than one). 
}
\label{fig:GaFeatureComp}
\end{figure*}

The top plots in Fig.~\ref{fig:GaFeatureComp} compare the reconstructed energy and Cherenkov angle of the T2K NCQE-like sample across different neutron capture multiplicities. 
The blue dashed line represents prompt events with zero detected neutrons, the red line corresponds to those with one neutron, and the orange dashed line depicts those with more than one neutron.
For events with zero neutrons, the reconstructed energy is generally lower, and the Cherenkov angle distribution has a taller peak around  $42^\circ$, which is consistent with light from a single $\gamma$.
Conversely, events with multiple detected neutrons tend to have higher reconstructed energies, leading to a concentration of multiple $\gamma$ events near $90^\circ$ in the Cherenkov angle distribution. 

The bottom six plots of Fig.~\ref{fig:GaFeatureComp} display the dataset alongside MC predictions, with results organized by neutron capture multiplicity. The second row corresponds to zero neutrons, the third row represents one neutron, and the bottom row illustrates cases with more than one neutron. Notably, the bottom right panel of Fig.~\ref{fig:GaFeatureComp} illustrates the Cherenkov angle distribution for events with more than one detected neutron. It is observed that the majority of data points cluster around approximately $42^\circ$ in the Cherenkov angle distribution, whereas MC predictions indicate that most events should concentrate around $90^\circ$. This discrepancy may be attributed to the limited statistics in this study. In the near future, the statistical sample could be enhanced by the data set collected during the pure water phase, corresponding to 14.94~×~10$^{20}$ POT, which is about 8.5 times the neutrino beam exposure used in this study. The statistics could also be improved by utilizing future data sets with a $\sim$0.03\% Gd concentration at SK~\cite{SK_2ndGdLoad}.

\subsection{Multiple scattering goodness}
The multiple scattering goodness (\textit{MSG}) is a reconstructed parameter recently introduced in the DSNB analysis flow at SK~\cite{Nu2024_DSNB}, though it was originally developed for the solar neutrino analysis~\cite{MSG_intro}.
This parameter is designed to assess the extent to which a particle has scattered during propagation in the detector by analyzing the pattern of PMT hits to determine their anisotropy.
At energies relevant to this analysis and the DSNB study at SK, lower energy electrons are more likely to undergo multiple Coulomb scattering than those at higher energies.
As a result, the lower energy particles typically produce a more isotropic pattern of hits after multiple scatterings. 
Since events with one or multiple $\gamma$'s have an even more isotropic pattern, \textit{MSG} is useful for separating the single-positron prompt event characteristic of the DSNB signal from NCQE backgrounds.
An \textit{MSG} value around 0.5 indicates consistency with a single particle direction, suggesting a more coherent Cherenkov pattern.
Lower values indicate more isotropy. 
The calculation of the \textit{MSG} is detailed in Ref.~\cite{MSG_intro}. 

\begin{figure*}[t]
\centering
\includegraphics[width=0.8\textwidth]{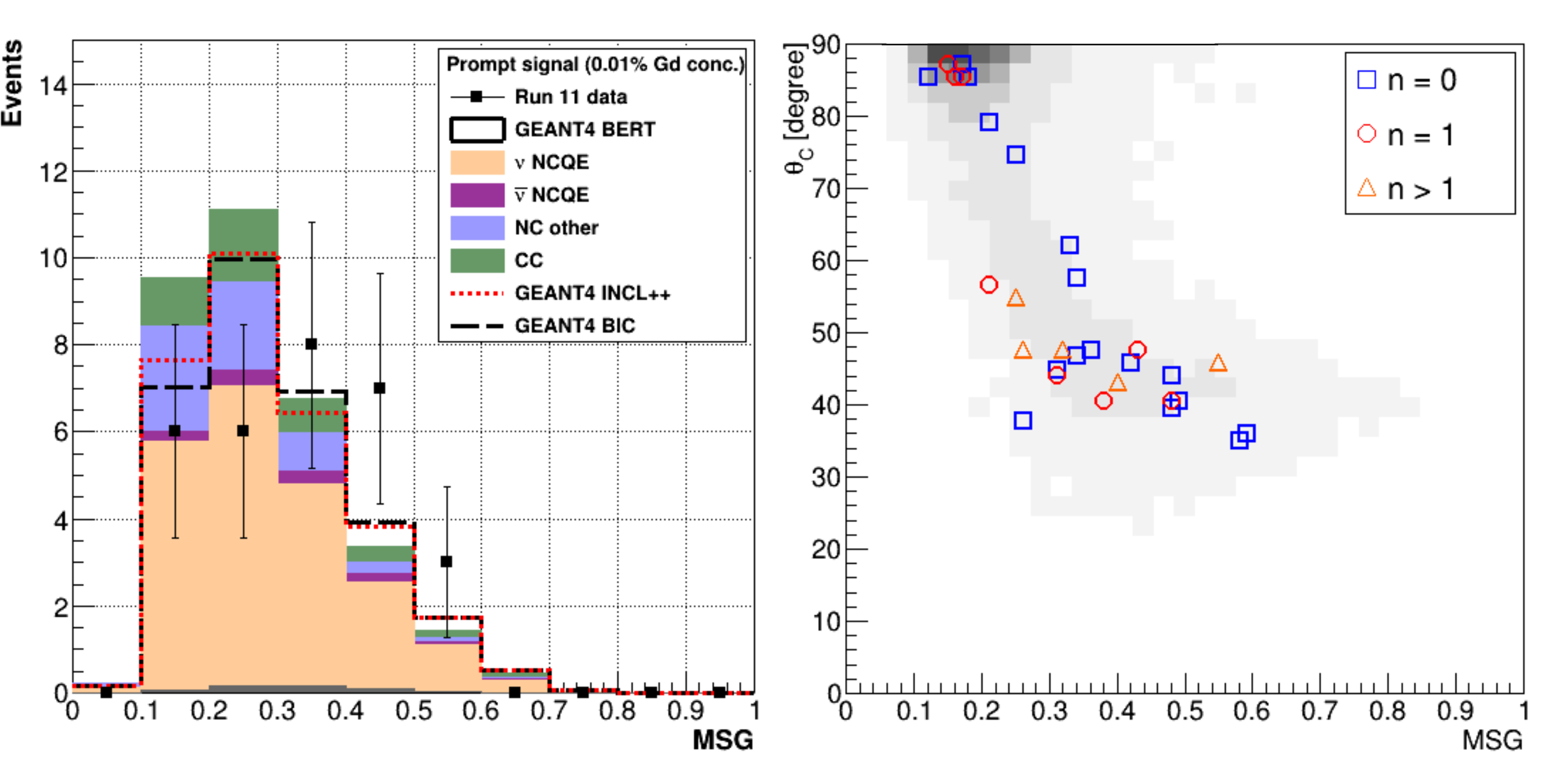}
\caption{For both panels, events in the full energy range [3.49, 29.49]~MeV are shown. Left panel: \textit{MSG} distributions for prompt events, comparing data set with MC using the BERT SI model (histogram), INCL++ (dotted red), and BIC (dashed black). Right panel: Correlation between Cherenkov angle and \textit{MSG}, with \textit{MSG} $\sim$ 0.5 indicating coherent patterns. Delayed signal multiplicity for each prompt event is categorized as zero (blue squares), one (red circles), or more than one (orange triangles).}
\label{fig:MSG_all}
\end{figure*}

The \textit{MSG} distribution of T2K NCQE-like events in Run 11, as shown in the left plot of Fig.~\ref{fig:MSG_all}. 
The histogram is generated using SKG4 with the BERT SI model, while the red dotted and black dashed lines correspond to the SI models INCL++ and BIC, respectively. 
The right plot of Fig.~\ref{fig:MSG_all} displays the $\theta_C$ versus \textit{MSG} distribution, where delayed signal multiplicities are labeled as follows: zero neutrons are marked with blue squares, one neutron with red circles, and more than one neutron with orange triangles. 
We note that the NCQE-like events tend to populate lower \textit{MSG} values as expected.
The distribution itself is a valuable tool for estimating uncertainties in the DSNB analysis stemming from this paramater.

\begin{figure*}[t]
\centering
\includegraphics[width=0.8\textwidth]{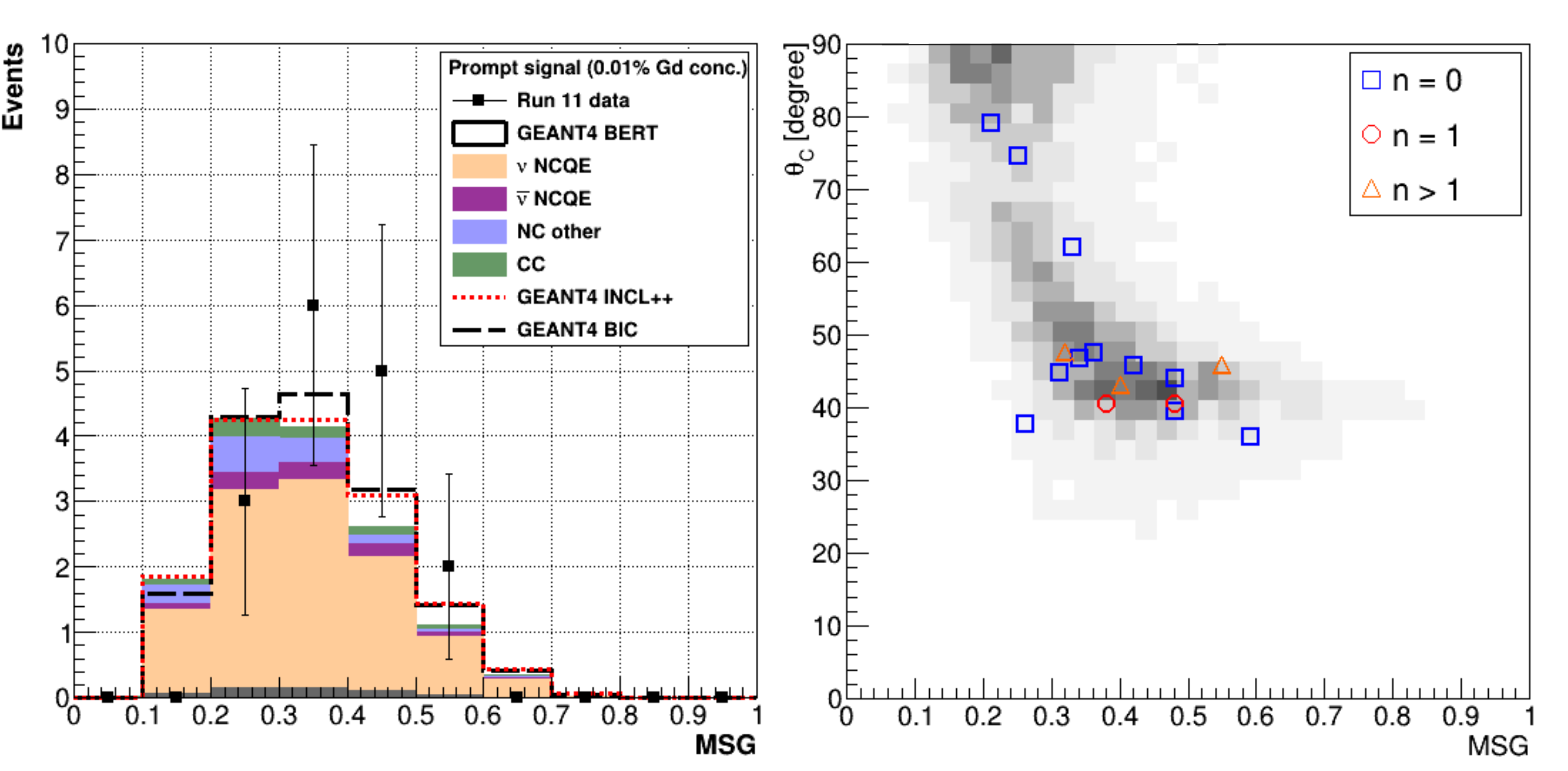}
\caption{Comparisons of data and MC predictions for NCQE-like events within the reconstructed energy range of [3.49, 7.49]~MeV, showing \textit{MSG} distributions alongside $\theta_C$.}
\label{fig:MSG_low}
\end{figure*}

\begin{figure*}[t]
\centering
\includegraphics[width=0.8\textwidth]{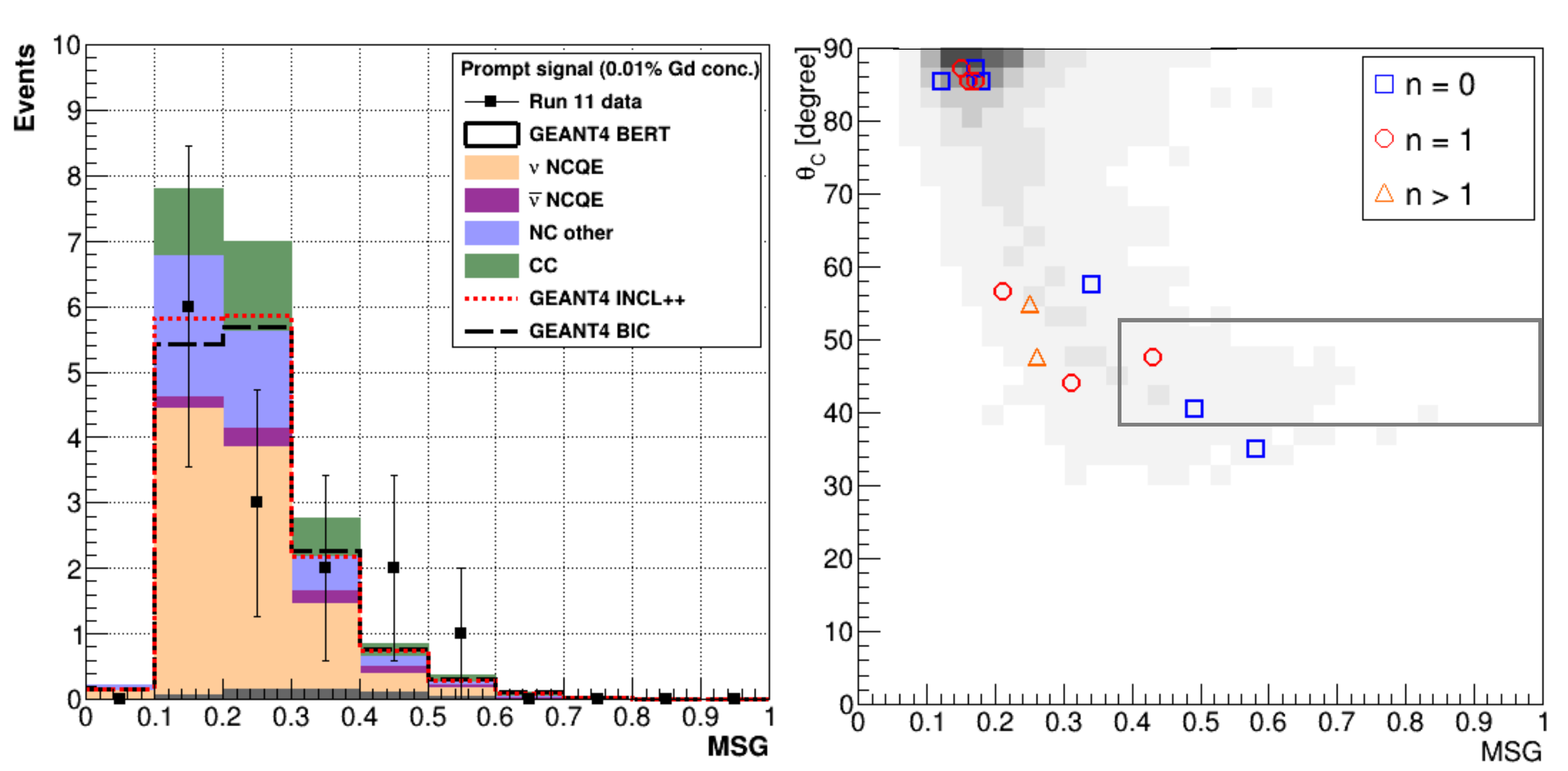}
\caption{Same as Fig.~\ref{fig:MSG_low}, but for the higher-energy range of [7.49, 29.49]~MeV, which overlaps with the DSNB search window. In the left plot, an \textit{MSG} > 0.38 selection yields an efficiency of 21.4\% $\pm$ 17.1 (stat.), while the right plot highlights (in gray) the naive DSNB cut region (38$^\circ$ < $\theta_C$ < 53$^\circ$, \textit{MSG} > 0.37).}
\label{fig:MSG_high}
\end{figure*}

The \textit{MSG} distribution varies with energy as is shown in Fig.~\ref{fig:MSG_low} (3.49 to 7.49~MeV) and Fig.~\ref{fig:MSG_high} (7.49 to 29.49 MeV). 
The agreement between the data and MC in both figures supports the use of \textit{MSG} information to exclude NCQE events in the DSNB study.
Fig.~\ref{fig:MSG_high} is particularly significant as it focuses on the reconstructed energy range that overlaps with the DSNB search. 
The efficiency of a \textit{MSG} $>$ 0.38 cut, shown in the left plot of Fig.~\ref{fig:MSG_high}, is 21.4\% ± 17.1\% (stat.).
In the right plot, the gray region highlights the naive cut settings for the DSNB search region at SK, defined as $38^\circ < \theta_C < 53^\circ$ with \textit{MSG} $>$ 0.37. 
However, we note that the cuts used in the SK DSNB search~\cite{Nu2024_DSNB} will be determined on an energy bin-by-bin basis. 
The information presented here will be useful for a more detailed assessment of the systematic uncertainties associated with the \textit{MSG} selection applied between atmospheric NCQE and DSNB events.

\nocite{*}

\bibliography{reference}

\end{document}